\documentclass[titlepage,11 pt]{article}

\usepackage[centertags]{amsmath}
\usepackage{graphicx}
\usepackage[english]{babel}
\usepackage{natbib}
\usepackage{url} 
\usepackage{pifont}
\usepackage{hyperref}
\usepackage{amsmath}
\usepackage{amsfonts}

\usepackage{multicol}
\usepackage{bbm}
\usepackage{url} 

\usepackage[affil-it]{authblk}

\usepackage{hyperref}

\usepackage{setspace}
\usepackage{lipsum}
\usepackage[margin=3cm]{geometry}

\numberwithin{equation}{section}
\newcommand{\Z}{\mathcal{Z}}
\newcommand{\matformat}{\mathbf}
\newcommand{\vecformat}{\mathbf}

\newtheorem{thm}{Theorem}[section]

\title{\textbf{A bootstrap test to detect prominent\\ Granger-causalities across frequencies}}

\author{\textbf{Matteo Farn\'{e}}  \thanks{Electronic address: \texttt{matteo.farne2@unibo.it}; Corresponding author}}
\affil{Department of Statistical Sciences,\\ University of
Bologna, Italy}

\author{\textbf{Angela Montanari}  
}
\affil{Department of Statistical Sciences,\\ University of
Bologna, Italy}

\begin{document}

\maketitle

\begin{abstract}
Granger-causality in the frequency domain is an emerging tool to analyze the causal relationship between two time series.
We propose a bootstrap test on unconditional and conditional Granger-causality spectra, as well as on their difference,
to catch particularly prominent causality cycles in relative terms.
In particular, we consider a stochastic process derived
applying independently the stationary bootstrap to the original series.
Our null hypothesis is that each causality or causality difference is equal to the median across frequencies computed on that process.
In this way, we are able to disambiguate causalities which depart significantly from the median one obtained ignoring the causality structure.
Our test shows power one as the process tends to non-stationarity, thus being more conservative than parametric alternatives.
As an example, we infer about the relationship between money stock and GDP in the Euro Area via our approach,
considering inflation, unemployment and interest rates as conditioning variables.
We point out that during the period 1999-2017 the money stock aggregate M1 had a significant impact on economic output
at all frequencies, while the opposite relationship is significant only at high frequencies.
\end{abstract}


%
%
%
%
%
%
%
%


\section{Introduction}

%
%


As a statistical concept, causality has a central role both from a theoretical and a practical point of view (see \cite{berzuini2012causality}).
In time series analysis, 
the concept that was to be called Granger-causality (GC) was first introduced by Wiener in the context
of prediction theory \citep{wiener1956theory} and then formalized by Granger in the
context of linear regression modelling of stochastic processes \citep{granger1969investigating}.
Causality measures in the frequency domain were first proposed in \cite{pierce1979r} as $R^2$ measures for time series.
In \cite{geweke1982measurement} and \cite{geweke1984measures} the fundamental concepts of unconditional and conditional
Granger-causality in the frequency domain were introduced (and extended
in \cite{hosoya1991decomposition} and \cite{hosoya2001elimination} respectively).

While the use of GC in the time-domain dates back to the sixties, GC in the frequency domain has become increasingly popular in recent years.
In \cite{lemmens2008measuring} the causality structure of European production expectation surveys is analyzed by the methods of \cite{pierce1979r} and
\cite{geweke1982measurement} comparatively, which require to study appropriate frequency-wise coefficients of coherence.
The same approach is used in \cite{tiwari2014frequency} for exploring the relationship between energy consumption and income in the United States.
The advantage of frequency-domain GC lies in the disentanglement of the causality structure across a range of frequencies,
while traditional time-domain GC only provides an overall indication on the presence of a causality relationship.
The aim of our paper is to provide an inferential tool to mark up the strongest causalities in the frequency domain,
in order to draw meaningful remarks about the causality structure.

In \cite{ding200617}, bootstrap thresholds are computed to make inference about Geweke's unconditional and conditional GC measures in the context of neurological data,
via the randomization approach of \cite{blair1993alternative}.
A further extension of that approach can be found in \cite{wen2013multivariate}, and relevant applications in the neurophysiological context include \cite{brovelli2004beta},
\cite{roebroeck2005mapping} and \cite{dhamala2008analyzing}, where explicit VAR estimation is avoided by a nonparametric approach.
More recently, a comprehensive computational and inferential strategy for time-domain and frequency-domain GC spectra
has been proposed in \cite{barnett2014mvgc}.



In \cite{breitung2006testing}, a parametric test for Granger-causality in the frequency domain is proposed.
Its convergence rate is $O(T^{-1/2})$ (where $T$ is the time length) and its power is decreasing as the
distance of the frequency of interest from $\frac{\pi}{2}$ increases (even if \cite{yamada2014some} show that the same test is still useful at extreme frequencies).
The test is based upon a set of linear restrictions on the parameters of the (possibly cointegrated) VAR model best representing the series (we refer to \cite{lutkepohl2005new} for VAR selection and estimation).

Applying such a test to time series with a rich causality structure, like macroeconomic ones,
most of causalities are often flagged as significant. 
In addition, test precision may suffer at extreme frequencies when $T$ is not large.
Nonetheless, such procedure is widely used in the literature.
For example, a relevant application for studying the relationship between real and financial business cycles can be found in \cite{gomez2015credit}.

Some nonparametric testing approaches were also proposed in the literature. \cite{hidalgo2000nonparametric} estimates VAR filters via generalized least squares and then accordingly derives a test statistics for GC. \cite{hidalgo2005bootstrap} extends the framework of \cite{hidalgo2000nonparametric} to the multivariate case computing relevant quantiles under the null via resampling bootstrap. 
In \cite{assenmacher2008interpeting}, the Philipps spectral estimator (see \cite{philipps1988spectral}) is exploited to estimate causality both at frequency $0$ and at the rest of frequencies in a cointegrated setting. Such a method is used in \cite{berger2011does} to test the relationship between money growth and inflation in the Euro Area.

The present work proposes a complementary approach to the classical testing framework of the no-causality hypothesis.
Our aim is to detect prominent cycles, i.e. cycles which are dominant compared to others for explaining the causality relationship.
We would like to answer the following question: 'Which causalities are larger than the median causality that would hold in case of stochastic independence?' The need for such a tool rises to distinguish the most relevant causalities for the causality structure of the process among significant causalities in the classical sense.

In order to reach this goal, we approximate the data generating process under the null applying the stationary bootstrap of \cite{politis1994stationary}
independently to each series.
We derive the desired bootstrap quantile of the median causality and we compare each causality to it.
The median is chosen because we need a unique comparison ground for each causality, and under the null there is no reason to suppose that causalities are stochastically different. Our test is adaptive with respect to the true spectral shape and can be used as a complementary tool to classical tests, provided that $T$ is large enough to ensure $T^{\frac{1}{3}} \rightarrow \infty$.

We exploit the described tool for studying the mutual relationship
between economic output and money stock in the Euro Area,
as, in so doing, we can identify characteristic frequencies,
i.e. characteristic time periods of the causality structure.
The problem has been widely addressed as far as the US economy is concerned.
In \cite{belongia2016money}, for instance, the methodology of \cite{friedman1975money},
based on structural VAR models, is revisited and
applied to U.S. data across the period $1967-2013$.
On the contrary, evidences and analyses regarding the Euro Area are still weak even if the
belief that money stock somehow affects business cycle is present in the
literature.

The paper is organized as follows. In the next Section the concept of
Granger-causality in the frequency domain is recalled, our bootstrap inference approach is explained in detail
and a simulation study which clarifies the features of our test is presented.
In Section \ref{sec3} we show the potentialities of our method in outlining
the causal relationships between money and output in the Euro Area
during the period 1999-2017.
Finally, we conclude the paper with a discussion.


\section{Granger-causality spectra: a bootstrap testing approach}\label{sec:G}

\subsection{Definition}
Let us suppose that the past values of a time series ${Y}_t$, i.e. ${Y}_{t-1}$, ${Y}_{t-2}$, $\dots$, help predicting the value at time $t$ of another time series ${X}_t$, that is,
${Y}_{t-1}$, ${Y}_{t-2}$, $\dots$ add significant information to
the past values of ${X}_t$ (${X}_{t-1}$, ${X}_{t-2}$, $\dots$) for predicting
${X}_t$. In that case we say that ${Y}_t$  Granger-causes ${X}_t$.

We now briefly recall the bases of Granger-causality spectral theory. We
follow the approach in \citet{ding200617}, which we refer to for the details.
Suppose that ${X}_t$ and ${Y}_t$, jointly covariance-stationary, follow a
non-singular $VAR(k)$ model. Defining $\vecformat{Z}_t=[{X}_t, {Y}_t]'$, we have
\begin{equation}\vecformat{Z}_t=\matformat{A}_{1}\vecformat{Z}_{t-1}+\ldots+\matformat{A}_{k}\vecformat{Z}_{t-k}+\vecformat{\epsilon}_t,\label{VAR}\end{equation}
where $\vecformat{\epsilon}_t\sim N_2(0,{\Sigma_2})$, ${\Sigma_2}$ is a $2 \times 2$ covariance matrix, and
$\matformat{A}_{1}, \ldots, \matformat{A}_{k}$ are $2 \times 2$ coefficient matrices.

Moving to the frequency domain, for each frequency $\omega$ we define the transfer function $\matformat{P}(\omega)$ of $\vecformat{Z}_t$ in (\ref{VAR}) as
\begin{equation}\matformat{P}(\omega)=\left(\matformat{I}-\sum_{j=1}^k \matformat{A}_j e^{-i j\omega}\right)^{-1}, \; -\pi \leq \omega \leq \pi,\label{transf}\end{equation}
which is invertible if and only if the roots of the equation $\det(\matformat{I}_p-\sum_{j=1}^{k}\matformat{A}_jL^j)~=~0$  (where $L$ is the lag operator)
lie within the unit circle.
Setting
$\matformat{P}(\omega)= $\\
$\left[\begin{array}{cc}
P_{{XX}}(\omega)& P_{{XY}}(\omega)\\
P_{{YX}}(\omega) & P_{{YY}}(\omega)
\end{array}\right]$,
the definition (\ref{transf}) allows to define in a compact way the model-based spectrum $h(\omega)$ as follows:
$$\matformat{h}(\omega)=\matformat{P}(\omega)\matformat{\Sigma_2} \matformat{P}(\omega)^{*},  \; -\pi \leq \omega \leq \pi,$$ where $^{*}$ denotes the complex conjugate.

Setting $\matformat{\Sigma_2}= \left[\begin{array}{cc}
\sigma_2& \upsilon_2\\
\upsilon_2 & \gamma_2
\end{array}\right]$,
we need for computational reasons to define the transform matrix $\matformat{S}=
\left[\begin{array}{cc}
1 & 0\\
-\frac{\upsilon_2}{\gamma_2} & 1
\end{array}\right],$
from which we derive the transformed transfer function matrix
$\matformat{\tilde{{P}}}(\omega)=\matformat{S} \times \matformat{P}(\omega)$. The
process $\vecformat{Z}_t=[{X}_t, {Y}_t]'$ is normalized accordingly as
$\matformat{Z}^{*}_t=\matformat{\tilde{{P}}}(L)[{X}_t,{Y}_t]'$ and becomes
$\matformat{Z}^{*}_t=[{X}^{*}_t, {Y}^{*}_t]'$.

The unconditional Granger-causality spectrum of ${X}_t$ (effect-variable) respect
to ${Y}_t$ (cause-variable) is then defined as \citep{geweke1982measurement}
\begin{equation}\label{uncond}
h_{{Y}\rightarrow X}(\omega)=\ln\left(\frac{{h}_{{XX}}(\omega)}{\tilde{{P}}_{{XX}}(\omega)\sigma_2\tilde{{P}}_{{XX}}(\omega)^{*}}\right).\end{equation}
In the empirical analysis, the theoretical values of coefficient and covariance matrices will be replaced by the corresponding SURE estimates \citep{zellner1962efficient}.

Moreover, we can define the conditional Granger causality spectrum of ${X}_t$
respect to ${Y}_t$ given an exogenous variable ${W}_t$ (conditioning variable).
Suppose we estimate a VAR on $[{X}_t, {W}_t]'$ with covariance matrix of the
noise terms $\matformat{\Sigma_{2'}}= \left[\begin{array}{cc}
\sigma_{2'}& \upsilon_{2'}\\
\upsilon_{2'} & \gamma_{2'}.
\end{array}\right]$ and
transfer function $\matformat{G}(\omega)$ (defined as in (\ref{transf})). The
corresponding normalized process of $[{X}_t, {W}_t]'$ (according to the procedure
described above) is denoted by $[{X}^{*}_t, W^{*}_t]'$.

We then estimate a VAR on $[{X}_t, {Y}_t, {W}_t]'$ with covariance matrix of the
noise terms
$$\matformat{{\Sigma}_3}=\left[\begin{array}{ccc}
\sigma_{{XX}} &\sigma_{{XY}}& \sigma_{{XW}} \\
\sigma_{{YX}} &\sigma_{{YY}}& \sigma_{{YW}}\\
\sigma_{{WX}} &\sigma_{{WY}}& \sigma_{{WW}} \end{array}\right]$$ and transfer
function $\matformat{P}'(\omega)$. Building the matrix
$$\matformat{C}(\omega)=\left[\begin{array}{ccc}
G_{{XX}}(\omega) &0& G_{{XW}}(\omega) \\
0 &1& 0\\
G_{{WX}}(\omega) &0& G_{{WW}}(\omega) \end{array}\right],$$ we can define
$\matformat{Q}(\omega)=\matformat{C}^{-1}(\omega) \matformat{P}'(\omega)$, which is a
sort of ``conditional'' transfer function matrix. The theoretical spectrum of
${X}^{*}$ can thus be written as
$$h_{{X}^{*}{X}^{*}}(\omega)=Q_{{XX}}(\omega)\sigma_{{XX}}
Q_{{XX}}(\omega)^{*}+$$
$$+Q_{{XY}}(\omega)\sigma_{{YY}}Q_{{XY}}(\omega)^{*}+Q_{{XW}}(\omega)\sigma_{{WW}}Q_{{XW}}(\omega)^{*}.$$
The conditional spectrum of ${X}_t$ (effect-variable) respect to ${Y}_t$
(cause-variable) given ${W}_t$ (conditioning variable) is \citep{geweke1984measures}
\begin{equation}\label{cond} h_{{Y}\rightarrow {X}|{W}}(\omega)=\ln\left(\frac{h_{{X}^{*}{X}^{*}}(\omega)}{Q_{{XX}}(\omega)\sigma_{{XX}}
Q_{{XX}}(\omega)^{*}}\right).\end{equation}

Both $h_{{Y}\rightarrow {X}}(\omega)$ and $h_{{Y}\rightarrow {X}|{W}}(\omega)$ range from $0$ to $\infty$, with $-\pi~\leq ~\omega ~\leq~\pi$. $h_{{Y}\rightarrow {X}}(\omega)$ expresses the power of the relationship
from ${Y}$ to ${X}$ at frequency $\omega$, $h_{{Y}\rightarrow {X}|{W}}(\omega)$ expresses
the strength of the relationship from ${Y}$ to ${X}$ at frequency $\omega$ given ${W}$. Therefore, the unconditional spectrum accounts for the whole
effect of the past values of ${Y}_t$ onto ${X}_t$, while the conditional spectrum accounts for the
direct effect of the past values of ${Y}_t$ onto ${X}_t$ excluding the effect mediated by the past values of ${W}_t$.
The same measures are more easily defined in the time-domain. In that case, they are
defined for the process as a whole (not frequency-wise as in the frequency
domain).

Granger-causality spectra $h_{{Y}\rightarrow {X}}(\omega)$ and $h_{{Y}\rightarrow {X}|W}(\omega)$
can be interpreted as follows.
If $h_{{Y}\rightarrow {X}}(\omega)>0$, it means that past values of ${Y}_t$ help
predicting ${X}_t$, and $\frac{1}{\omega}$ is a relevant cycle. If
$h_{{Y}\rightarrow {X}|W}(\omega)>0$, it means that past values of ${Y}_t$ in
addition to those of ${W}_t$ help predicting ${X}_t$, and $\frac{1}{\omega}$ is a
relevant cycle. Significant frequencies give us some hints on the relevant
delay structure of the cause variable with respect to the effect variable.

We remark that these measures do not give any information on the sign of the
relationship, which is given by time-domain measures like the correlation
coefficient. It rather describes the strength, i.e. the intensity, of the
causal relationship.

\subsection{Testing framework}\label{sec:frame}
The inference on Granger-causality spectra in the frequency
domain is still an open problem. In fact, differently from the corresponding time-domain quantities,
the limiting distribution for unconditional and conditional spectra is
unknown (see \citet{barnett2014mvgc}, section 2.5). In spite of that,\\ \cite{breitung2006testing} test
the nullity of unconditional and conditional GC at each frequency $\omega$, imposing
a necessary and sufficient set of linear restrictions to the (possibly cointegrated) VAR model best fitting the series.
The resulting test statistics is distributed under the null as a Fisher distribution with $2$ and $T-2k$ degrees of freedom
(except for $\omega=\{0,\pi\}$ at which the distribution is $F_{(1,T-k)}$), where $k$ is the VAR delay and $T$ is the time series length.

As said in the Introduction, that test applied on macroeconomic series often flags most of causalities as significant, due to the rich causality structure.
For this reason, in order to disambiguate among significant causalities the most prominent ones, we propose a complementary bootstrap testing approach.
At each frequency $\omega$, we test the \textit{null} hypothesis $H_0: t(\omega)=t_{med}$, against the \textit{alternative} $H_1: t(\omega)>t_{med}$, where the functional $t(\omega)$ may be the unconditional GC $h_{{Y}\rightarrow X}(\omega)$, the conditional GC $h_{{Y}\rightarrow {X}|{W}}(\omega)$ or their difference $h_{{Y}\rightarrow X}(\omega)-h_{{Y}\rightarrow {X}|{W}}(\omega)$, and $t_{med}$ is the median of $t(\omega)$ across frequencies under the assumption of stochastic independence.




Since the distributions of $h_{{Y}\rightarrow X}(\omega)$ and $h_{{Y}\rightarrow {X}|{W}}(\omega)$ are unknown,
we approximate the distribution of each $t(\omega)$ under the null by the \textit{stationary bootstrap} of \cite{politis1994stationary}.
A similar approach was originally proposed by \cite{ding200617}, 
which tests the same null hypothesis of \cite{breitung2006testing} by the randomization procedure of \cite{blair1993alternative},
retaining the maximum causality across frequencies. Differently,
our procedure tests by bootstrap the equality between each unconditional or conditional causality and the median causality
under the assumption of stochastic independence.
Applying the stationary bootstrap to each time series independently of the other ones
approximates the no-causality situation,
because it approximates the Markov chain best representing independently each series.
We stress that unconditional and conditional spectra must be assessed separately, because their distributions are in general different.

In \cite{ding200617}, the comparison between unconditional and conditional
Granger causalities is performed using the randomized t-test of \cite{blair1993alternative}
on the bootstrapped series
of the causality peak across frequencies. This approach is suitable for their case,
where they perform psychological/neurological experiments, which allow to have multiple trials data.
On the contrary, in the
economic context, we can not perform such a test because we only have a single realization.
This is the reason why we take the difference between unconditional and conditional GC, which allows to
determine if the conditioning variable has a significant impact (amplification or annihilation)
on the causal relationship in our time-dependent data context.


Our idea derives from \cite{politis1994stationary}, according to which
each \textit{Fr\'{e}chet-differentiable} functional may be successfully approximated by the stationary bootstrap,
and the resulting bias depends on the sum of the \textit{Fr\'{e}chet-differential} $h_F$ evaluated at each observation,
given that the distance between the empirical and the true distribution function of $X_t$ is small.

The bootstrap series obtained via the stationary bootstrap of\\ \cite{politis1994stationary} are stationary Markov chains conditionally on the data.
It means that each bootstrap series $X_1^{*},\ldots,X_T^{*}$ is a Markov chain conditionally on $X_1,\ldots,X_T$.
Suppose that we apply the same procedure to $X_t$, $Y_t$ and $W_t$, obtaining the stationary bootstrap series $X_t^{*}$, $Y_t^{*}$ and $W_t^{*}$.
Computing unconditional and conditional Granger causality spectra on those series equals to assess causalities under the assumption of stochastic independence,
because the entire stochastic behaviours of $X_t^{*}$, $Y_t^{*}$ and $W_t^{*}$ are explained by the conditional distributions of
$X_t^{*}|X_{t-1}^{*}$, $Y_t^{*}|Y_{t-1}^{*}$, $W_t^{*}|W_{t-1}^{*}$.
Therefore, testing each Granger-causality computed on the original series $X_t$, $Y_t$ and $W_t$ against the median
causality computed across frequencies on $X_t^{*}$, $Y_t^{*}$ and $W_t^{*}$
is effective as a test for causality strength in relative terms.

Let us consider $\hat{r}(\omega)=\hat{h}_{{Y}\rightarrow X}(\omega)$,
which is defined as (\ref{uncond}) where the coefficient matrices $\matformat{A}_j, j=1,\ldots,k$, and the error covariance matrix $\matformat{\Sigma_2}$
are replaced by the corresponding SURE estimates \citep{zellner1962efficient}.
We know that SURE estimates $\matformat{\hat{A}}_j$, $j=1,\ldots,k$, are rational functions of the data, thus being \textit{Fr\'{e}chet-differentiable}.
$\matformat{\hat{\tilde{{P}}}}(\omega)$ and $\matformat{\hat{\Sigma}}_2$ are functions of the $\matformat{\hat{A}}_j$, thus being rational in turn; the same holds as a consequence for $\hat{{{h}}}(\omega)$.
Therefore, $\hat{h}_{{Y}\rightarrow X}(\omega)$, the natural logarithm of a rational function of the data, is \textit{Fr\'{e}chet-differentiable}.
At this point, as pointed out in \cite{politis2012subsampling}, page 30, even if the median is not \textit{Fr\'{e}chet-differentiable},
the bootstrap for $\hat{h}_{{Y}\rightarrow X}(\omega)$ is still valid, provided that the density function of $median({h}_{{Y}\rightarrow X}(\omega))$ is positive.
As a consequence, according to \cite{politis1994stationary}, paragraph 4.3, we can estimate consistently any quantile of the distribution
of the median of $\hat{h}_{{Y}\rightarrow X}(\omega)$ under the null hypothesis.

Considering $\hat{r}(\omega)=\hat{h}_{{Y}\rightarrow {X}|{W}}(\omega)$,
which is defined as (\ref{cond}) where the coefficient matrices and the error covariance matrix
are replaced by the corresponding SURE estimates, a similar reasoning can be carried out.
The same applies to the estimated difference of $\hat{h}_{{Y}\rightarrow X}(\omega)-\hat{h}_{{Y}\rightarrow {X}|{W}}(\omega)$, that is
\textit{Fr\'{e}chet-differentiable} apart from the case
$\hat{h}_{{Y}\rightarrow X}(\omega)=\hat{h}_{{Y}\rightarrow {X}|{W}}(\omega)$, which holds with null probability.

We stress that our aim is not to represent the common multivariate distribution function $F$
of the process $\mathbf{Z}_t=[X_t, Y_t, W_t]$.
That problem is an estimation one, which would be effectively solved by parametric or residual bootstrap.
Our aim is to exploit the random process $\mathbf{Z}_t^{*}=[X_t^{*}, Y_t^{*}, W_t^{*}]$ to derive the bootstrap quantile $q_{1-\alpha}^{*}$
which satisfies $$P(r_{med}^{*}\leq q_{1-\alpha}^{*})=1-\alpha,$$
where $\alpha$ is the significance level and $r_{med}^{*}$ is the bootstrap median across frequencies
of unconditional, conditional GC or their difference under the assumption of stochastic independence.
Since $r_{med}$ is \textit{Fr\'{e}chet-differentiable}, $P(r_{med}^{*}\leq q_{1-\alpha}^{*})$ approximates consistently
$P(r_{med}\leq q_{r,1-\alpha})$ as $T^{\frac{1}{3}}\rightarrow \infty$ under the null hypothesis of stochastic independence.

In more detail, suppose that $r$ is a \textit{Fr\'{e}chet-differentiable} functional, that is, there exists some influence function $h_F$
such that $$r(G)=r(F)+\int h_F d(G-F)+o(||G-F||)$$ with $\int h_F dF=0$ ($||.||$ is the supremum norm).
We define the mixing coefficient $$\alpha_{X}(k)=\sup_{A,B} |P(A,B)-P(A)P(B)|$$ where $A$ and $B$ vary over events in the $\sigma$-fields generated by $\{X_t, t \leq 0\}$ $\{X_t, t \geq k\}$.
The following Theorem holds.
\begin{thm}\label{appr_all}
Suppose that $X_t$, $Y_t$ and $W_t$ are strictly stationary random variables with distribution functions $F_{X}$, $F_{Y}$, $F_{W}$.
Assume that, for some $d \geq 0$, $E(h_{F_X}(X_{1})^{2+d})<\infty$, $\sum_{k} \alpha_{X}(k)^{\frac{d}{2+d}}<\infty$ and $\sum_{k} k^2 \alpha_{X}(k)^{1/2-\tau}< \infty$.
Further assume that these assumptions also hold for $Y_t$ and $W_t$. Then, if the distribution function $F_{\vecformat{Z}}$ of the random vector $\vecformat{Z}_t=[X_t, Y_t, W_t]$ can be factorized as $F_{X}F_{Y}F_{W}$, it holds
$$P(r(\hat{F}^{*}_{\vecformat{Z}})-r(F_{\vecformat{Z}})\leq q_r(1-\alpha))=1-\alpha$$
for any \textit{Fr\'{e}chet-differentiable} functional $r$ under the assumption $T^{\frac{1}{3}} \rightarrow \infty$.
\end{thm}
We refer to Appendix \ref{proofs} for the proof.

Due to the nature of our test, we need to exclude any stochastic process
with a constant Granger-causality spectrum different from a white noise.
Suppose that $\mathbf{Z}_t$ is a stochastic process with auto-covariance matrices $\mathbf{R}_j=cov(\mathbf{Z}_t,\mathbf{Z}_{t-j})$, $j \in \Z^+$.
Throughout the paper, we need
to assume that, in case there is at least \textit{one non-zero} causality coefficient at one delay $j \in \Z^+$, 
the resulting covariance matrix is \textit{not} diagonal, i.e. the effect and the cause variable are \textit{not} uncorrelated.
At the same time, we need to assume that each auto-covariance matrix $R_j$, $j\geq 1$, is positive definite.

We clarify the expressed constraints by an example.
Consider the case of a $VAR(1)$ with the following parameters: $\matformat{\Sigma}=diag(1,1)$,
$\matformat{A}_1=\left[\begin{array}{cc}
0& 0.5\\
0 & 0
\end{array}\right]$. According to \cite{wei2006time}, p. 392, the vectorized covariance matrix of a vector $AR(1)$ process is
$vec(R_0)=(\matformat{I}-\matformat{A}_1\otimes \matformat{A}_1)^{-1}vec(\matformat{\Sigma})$ and the vectorized autocovariance matrix at lag $k$ is $vec(\matformat{R}_k)=vec(\matformat{R}_0)\matformat{A}_1^k$.
Therefore, we have $\matformat{R}_0=diag(1.25,1)$, while $\matformat{R}_1$, $\matformat{R}_2$, $\ldots$ are singular matrices.
The same holds, for instance, if we suppose $\matformat{A}_1=\left[\begin{array}{cc}
0.5& 0.5\\
0 & 0
\end{array}\right]$. In that case, the covariance matrix results $\matformat{R}_0=diag(\frac{5}{3},1)$.

Such cases cannot be dealt by our procedure,
as SURE estimates are inconsistent. 
Out of this pathological set, our test achieves a power of $100\%$ if the underlying process is non-stationary.
On the contrary, the test of\\ \cite{breitung2006testing} is less conservative,
flagging surely as significant any causality distant enough from $0$.


\subsection{Testing procedure}

We now report in detail the testing procedures relative to the three functionals.
\begin{itemize}
\item
For the functional $\hat{h}_{{Y}\rightarrow X}(\omega)$, our bootstrap procedure is
\begin{itemize}{}
\item Simulate $N$ stationary bootstrap series $(X^{*}_t,Y^{*}_t)$ given the observed series $(X_t,Y_t)$. 
\item On each simulated series $(X^{*}_t,Y^{*}_t)$:
\begin{enumerate}
\item estimate a VAR model on $(X^{*}_t,Y^{*}_t)$ via SURE using BIC for model selection.
\item at Fourier frequencies $f_i=\frac{i}{T}, i=1,\ldots,[\frac{T}{2}]$,
compute $h_{{Y^{*}}\rightarrow X^{*}}(2\pi f_i)$. 
\item compute $median_{\{f_i, i=1,\ldots,T/2\}}h_{{Y^{*}}\rightarrow X^{*}}(2\pi f_i)$.
\end{enumerate}
\item Then, compute $q_{uncond,1-\alpha}$, the $(1-\alpha)$-quantile of the bootstrap distribution at Step 3 across the $N$ bootstrap series,
where $\alpha$ is the significance level.
\item Finally, at each $f_i$, \textit{flag} $\hat{h}_{{Y}\rightarrow X}(2\pi f_i)$ as \textit{significant} if larger than $q_{uncond,\alpha}$.
\end{itemize}
\item
For the functional $\hat{h}_{{Y}\rightarrow {X}|{W}}(\omega)$, the procedure becomes
\begin{itemize}
\item Simulate $N$ stationary bootstrap series $(X^{*}_t,Y^{*}_t,W^{*}_t)$ given the observed series $(X_t,Y_t,W_t)$. 
\item On each simulated series $(X^{*}_t,Y^{*}_t,W^{*}_t)$:
\begin{enumerate}
\item estimate a VAR model on $(X^{*}_t,W_t^{*})$ and $(X^{*}_t,Y^{*}_t,W^{*}_t)$ via SURE using BIC for model selection.
\item at Fourier frequencies $f_i=\frac{i}{T}, i=1,\ldots,[\frac{T}{2}]$,
compute $h_{{Y^{*}}\rightarrow {X^{*}}|{W^{*}}}(2\pi f_i)$. 
\item compute $median_{\{f_i, i=1,\ldots,T/2\}}h_{{Y^{*}}\rightarrow {X^{*}}|{W^{*}}}(2\pi f_i)$.
\end{enumerate}
\item Then, compute $q_{cond,1-\alpha}$, the $(1-\alpha)$-quantile of the bootstrap distribution at Step 3 across the $N$ bootstrap series.
\item Finally, at each $f_i$, \textit{flag} $\hat{h}_{{Y}\rightarrow {X}|{W}}(2\pi f_i)$ as \textit{significant} if larger than $q_{cond,\alpha}$.
\end{itemize}
\item
For the functional $\hat{h}_{{Y}\rightarrow X}(\omega)-\hat{h}_{{Y}\rightarrow {X}|{W}}(\omega)$, the procedure is
\begin{itemize}
\item Simulate $N$ stationary bootstrap series $(X^{*}_t,Y^{*}_t,W^{*}_t)$ given the observed series $(X_t,Y_t,W_t)$. 
\item On each simulated series $(X^{*}_t,Y^{*}_t,W^{*}_t)$:
\begin{enumerate}
\item estimate a VAR model on $(X^{*}_t,Y^{*}_t)$, $(X^{*}_t,W^{*}_t)$ and $(X^{*}_t,Y^{*}_t,W^{*}_t)$ via SURE using BIC for model selection.
\item at Fourier frequencies $f_i=\frac{i}{T}, i=1,\ldots,[\frac{T}{2}]$, compute
$h_{{Y^{*}}\rightarrow X^{*}}(2\pi f_i)-h_{{Y^{*}}\rightarrow {X^{*}}|{W^{*}}}(2\pi f_i)$. 
\item compute $median_{\{f_i, i=1,\ldots,T/2\}}h_{{Y^{*}}\rightarrow X^{*}}(2\pi f_i)-
h_{{Y^{*}}\rightarrow {X^{*}}|{W^{*}}}(2\pi f_i)$.
\end{enumerate}
\item Then, compute $q_{diff,\frac{\alpha}{2}}$ and $q_{diff,1-\frac{\alpha}{2}}$,
the $(\frac{\alpha}{2})$- and $(1-\frac{\alpha}{2})$-quantiles of the bootstrap distribution at Step 3 across the $N$ bootstrap series.
\item Finally, at each $f_i$, \textit{flag} $\hat{h}_{{Y}\rightarrow X}(2\pi f_i)-\hat{h}_{{Y}\rightarrow {X}|{W}}(2\pi f_i)$
as \textit{significant} if smaller than $q_{diff,\frac{\alpha}{2}}$ or larger than $q_{diff,1-\frac{\alpha}{2}}$.
\end{itemize}
\end{itemize}
We provide an R package, called ``grangers'', which performs these routines.

In addition, we extend our framework to test the nullity of $r(2\pi f_i)$, $i=1,\ldots,[\frac{T}{2}]$, across the frequency range.
In order to do that, we apply Bonferroni correction, that is, we apply the test procedure to each frequency with significance level $\frac{2\alpha}{T}$.
In this way, we ensure that the overall level is not larger than $\alpha$ under the null.
This approach is conservative: anyway, the test still has a power of $100\%$ as the VAR process tends to non-stationarity.

\subsection{Test features and simulation results}

In order to clarify the interpretation of our results,
we need to define the concept of ``prominence'' in a formal way.
At a \textit{significance level} $\alpha$, given a random time series sampled by the underlying data generating process,
any functional $r(\omega)$ is said to be maximally prominent at frequency $\omega$ if
$P\{r(\omega)>r_{med}\}> 1-\alpha$,
where $r_{med}$ is the median of $r(\omega)$ across frequencies.
As a consequence, the \textit{power} of our test procedure approaches $1$ as $r(\omega)$ is \textit{maximally prominent}.


We define the prominence rate at frequency $\omega$ as the expected probability of $r(\omega)$ to be maximally prominent:
$prom(\omega)$: $P(r(\omega)>q_{1-\alpha})$.
The prominence rate answers the question ``Which is the probability that $r(\omega)$ is maximally prominent?''
The degree of prominence at frequency $\omega$ is then defined as $dp(\omega)$: $P(r(\omega)>r_{med})$.
Instead, the power at frequency $\omega$ is defined as $power(\omega)=P(\hat{r}(\omega)>q_{1-\alpha})$.
Denoting the solutions of the characteristic equation $\det(\matformat{I}_p-\sum_{j=1}^{k}\matformat{A}_jL^j)~=~0$ in decreasing order by $\lambda_1,\ldots,\lambda_q$,
the maximal power at frequency $\omega$ is defined as $mp(\omega)=\lim_{|\lambda_1| \rightarrow 1} power(\omega)$.
For our test we observe $\max_{\omega \in ]0,2\pi]} mp(\omega)=1$.

In general, $\hat{r}(\omega)$ is \textit{significant} if larger than ${r}_{med}^{*}$ at a significance level $\alpha$.
As explained in Section \ref{sec:frame}, the distribution of ${r}_{med}^{*}$ consistently resembles the one of ${r}_{med}$ by the stationary bootstrap.
The \textit{level} of our test, as expected, is approximately equal to the chosen significance level $\alpha$
under the null, \textit{i.e.} in case of zero-causality at all frequencies (white noise process).

%

We now describe the performance of our test in a number of situations.
First of all, suppose that we simulate $100$ replicates from a VAR process in the form (\ref{VAR}) with $k=1$, $\matformat{\Sigma}=diag(1,1)$ and no causality coefficients.
The VAR delay is selected for each bootstrap setting by BIC criterion.
Our tested coefficient matrix $\matformat{A}_1$ is ${A_{1,(jj)}}=0,0.2,0.5,0.8,1$, $j=1,2$.
We observe that the estimated level is below $5\%$ at all Fourier frequencies, as long as $A_{1,(jj)}$ is distant from $1$.
If $A_{1,(jj)}=1$ (double random walk), the rejection rate increases at low frequencies,
according to the shape of prominence rate and degree of prominence, until $0.4$.

\begin{figure}[htbp]
\centering
\makebox{
\includegraphics[width=3in]{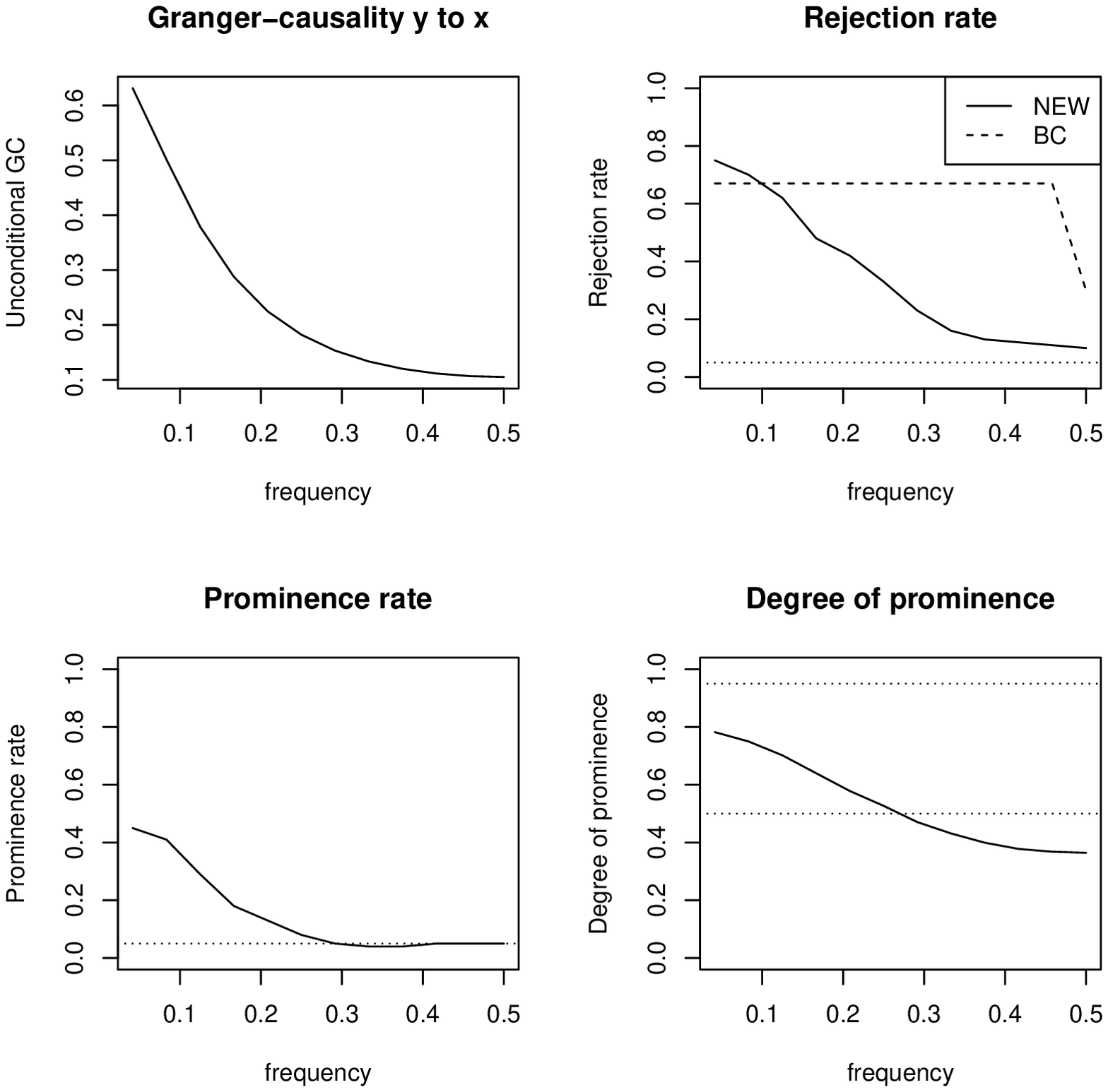}}
\caption{Case with $k=1$, ${A_{1,(j2)}}=0.5$, $j=1,2$. In dotted the significance level $\alpha=0.05$ and
the neutral degrees of prominence $0.5$ and $0.95$.  In dashed the rejection rate of BC test.} \label{uncond_zeros_again_0_5_top}
\centering
\makebox{
\includegraphics[width=3in]{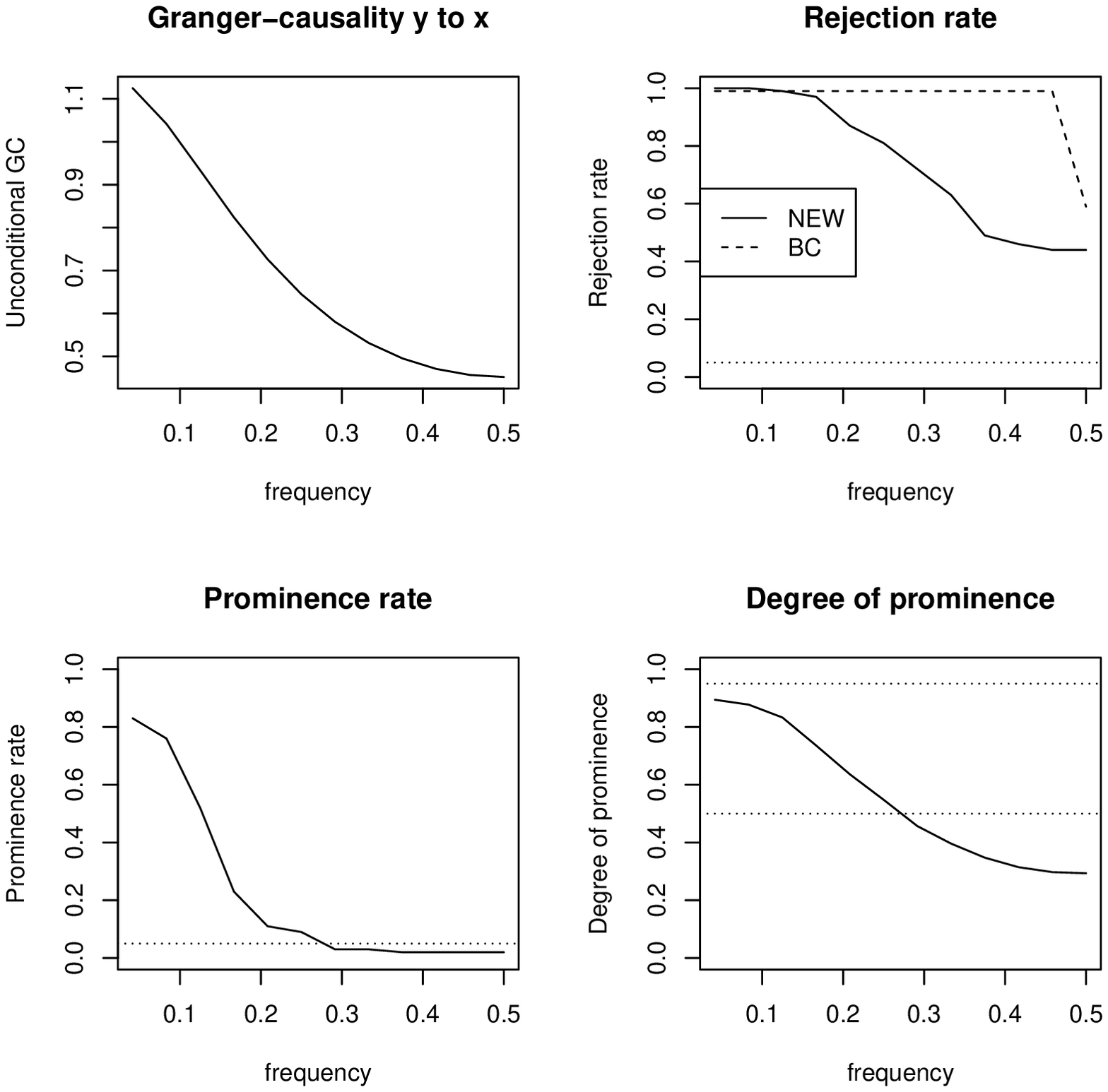}}
\caption{Case with $k=1$, ${A_{1,(j2)}}=1$, $j=1,2$.} \label{uncond_zeros_again_1_top}
\end{figure}

Another relevant case we deal with is for $k=1$ and ${A_{1,(j2)}}=0.5,1$, $j=1,2$. This process has an unconditional causality which decreases as the frequency increases.
For ${A_{1,(j2)}}=0.5$ (Figure \ref{uncond_zeros_again_0_5_top}), the degree of prominence ranges from $0.8$ to $0.3$ and the rejection rate is above $5\%$ at all frequencies, ranging from $0.9$ to $0.3$ approximately.
For ${A_{1,(j2)}}=1$ (Figure \ref{uncond_zeros_again_1_top}), the power at the lowest frequency is one, reflecting the prominence rate and the degree of prominence.
The same case is tested for the conditional causality, with very similar results.

\begin{figure}[htbp]
\centering
\makebox{
\includegraphics[width=3in]{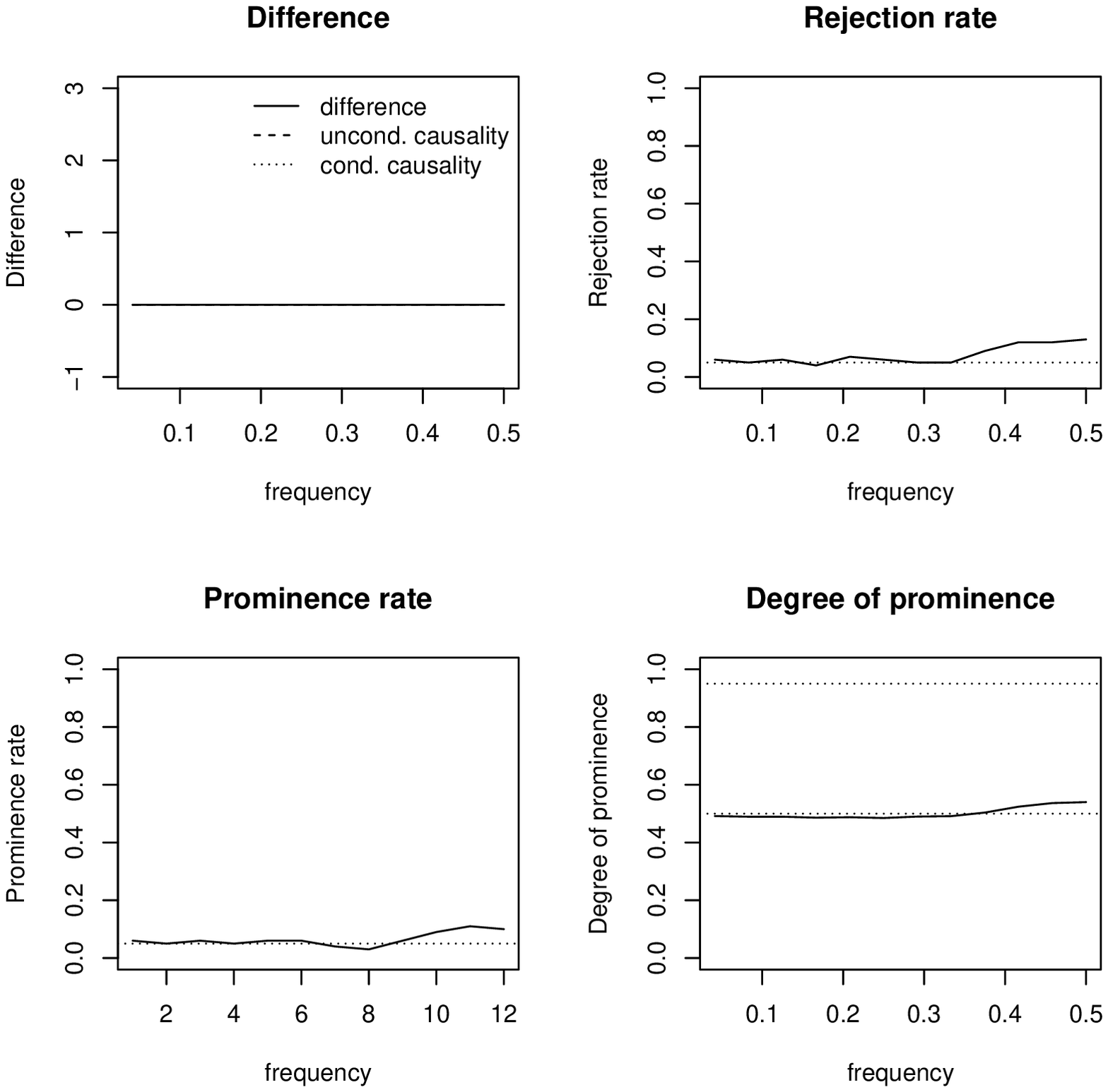}}
\caption{Comparing an unconditional and a conditional zero causality.} \label{zeros_again_diff_all}
\centering
\makebox{
\includegraphics[width=3in]{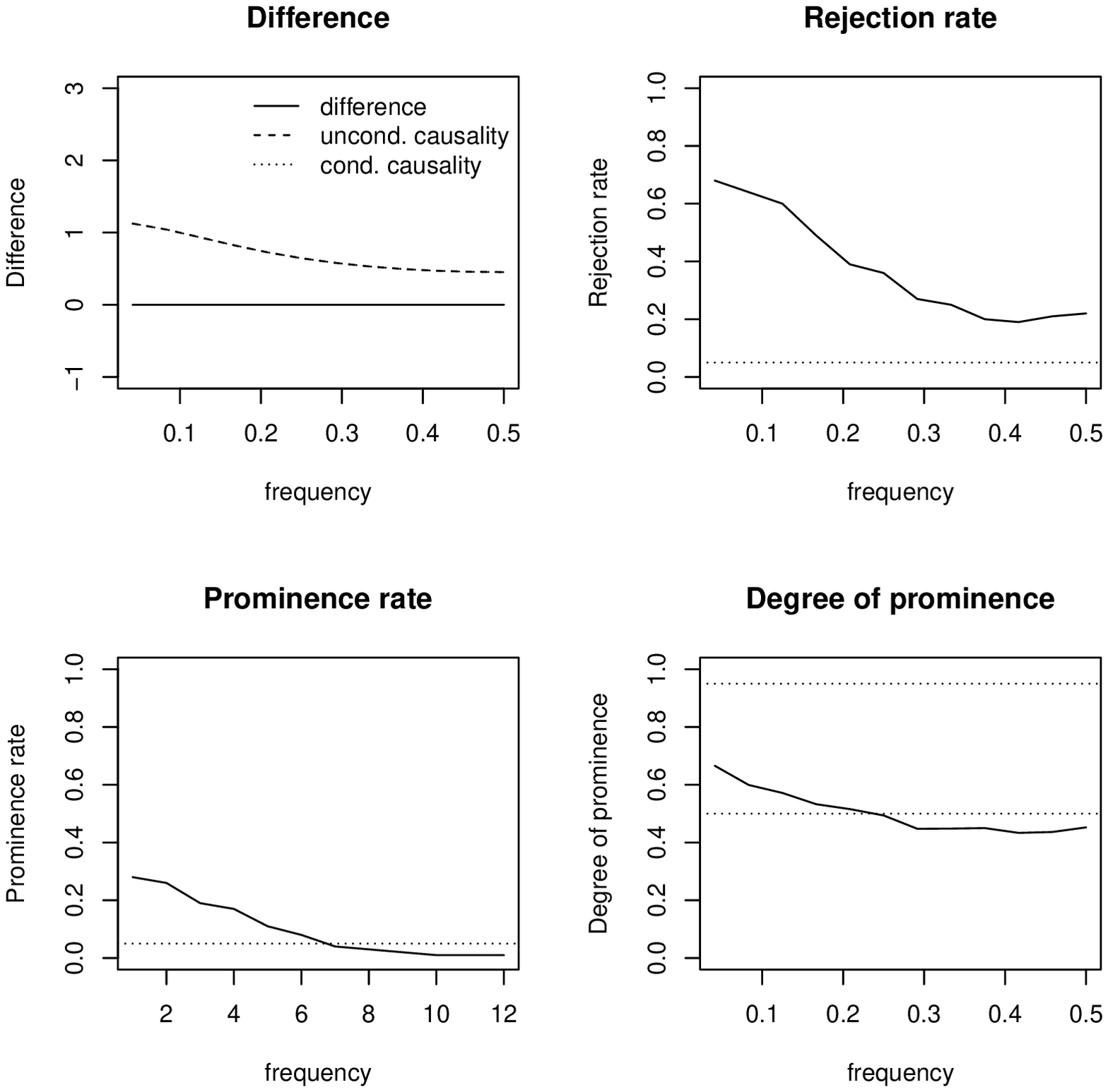}}
\caption{Comparing an unconditional and a conditional decreasing causality ${A_{1,(j2)}}=1$.} \label{decresc1top_diff_true_yes}
\end{figure}


We now compare an unconditional and a conditional causality which are zero at all frequencies.
For both cases, the rejection rates stand below $5\%$ at all frequencies.
If we compare two decreasing causalities having the shape above described (${A_{1,(j2)}}=0.5,1$, $j=1,2$), the rejection rate at the lowest frequency tends to increase until $0.4$ if the non-zero coefficients are equal to $1$ (the limit case of a double random-walk).
This pattern reflects the shape of prominence rate and degree of prominence (Figure \ref{decresc1top_diff_true_yes}).

If we compare an unconditional null causality to a decreasing conditional causality with parameters $0.5$ and $1$, the rejection rate is above $5\%$ at all frequencies and increases to $0.6$ or to $1$ respectively at the lowest frequency. In the latter case, we are in presence of a maximally prominent causality difference, as the degree of prominence and the prominence rate confirm.

Moreover, consider the VAR models described in \cite{breitung2006testing}, paragraph 4.
Those models have $k=3$, ${A_{k,(j2)}}=1$, $k=1,3$, $j=1$ and ${A_{k,(j2)}}=-2\cos(\omega^{*})$, $k=2$, $j=1$.
Such coefficient structure results in a null causality at frequency $\omega^{*}$.
On these settings, we can compare the results of our test to the results of ``BC test'' by \cite{breitung2006testing}, which appear in dashed line.
In addition, we test the sensitivity of the results to the condition number of the covariance matrix, setting $\matformat{\Sigma}=diag(1,1)$, $\matformat{\Sigma}=diag(0.2,1)$, $\matformat{\Sigma}=diag(5,1)$.

\begin{figure}[htbp]
\centering
\makebox{
\includegraphics[width=3in]{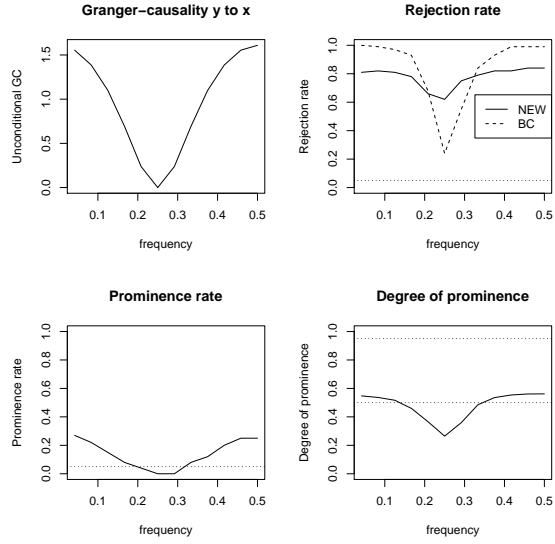}}
\caption{Case with $\omega^{*}=\frac{\pi}{2}$, $A_{k,(22)}=0$, $\matformat{\Sigma}=diag(1,1)$, $k=1,3$. In dashed the rejection rate of BC test.} \label{pimezzi_top}
\end{figure}

\begin{figure}[htbp]
\centering
\makebox{
\includegraphics[width=3in]{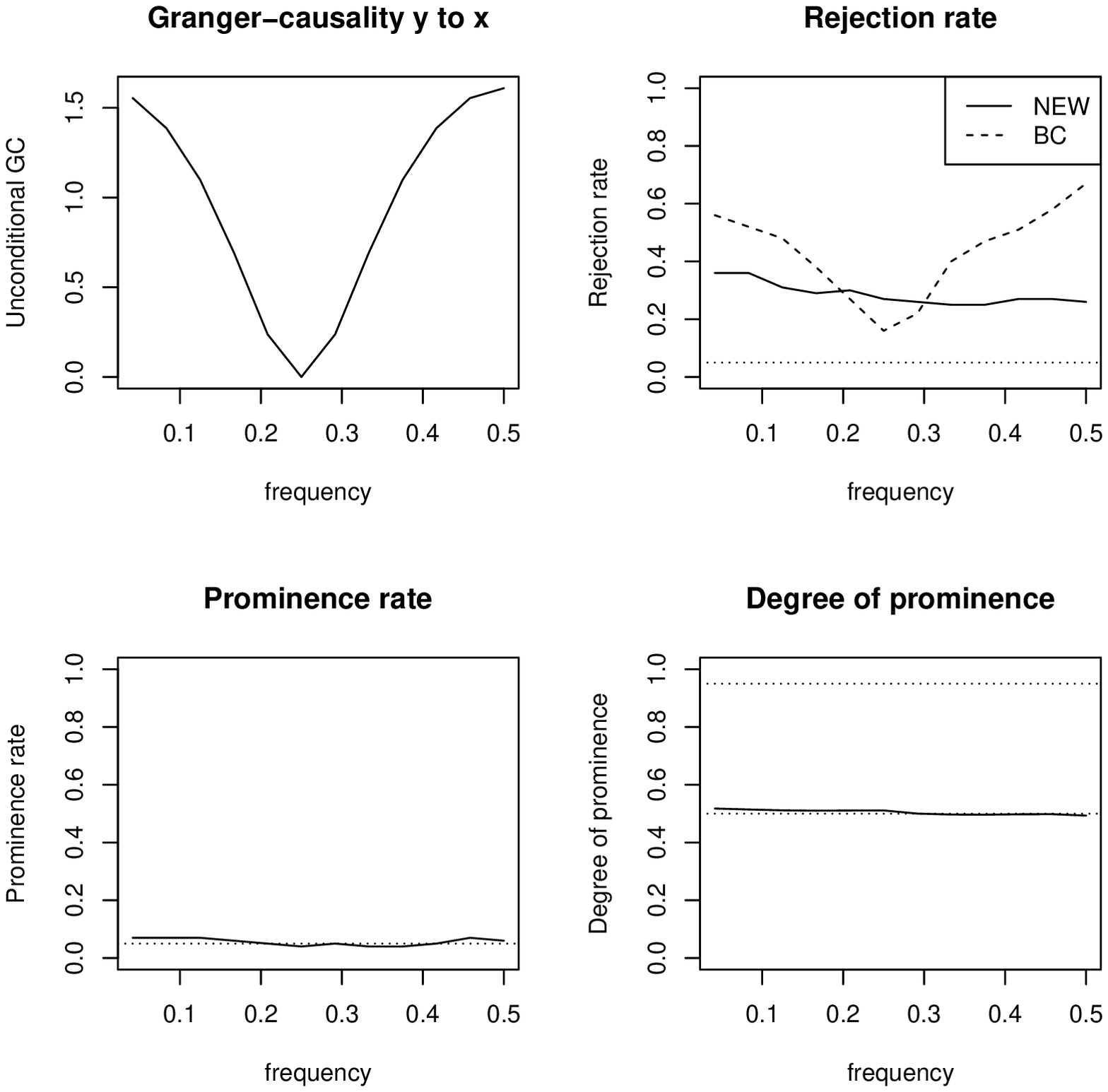}}
\caption{Case with $\omega^{*}=\frac{\pi}{2}$, $A_{k,(22)}=0$, $\matformat{\Sigma}=diag(0.2,1)$, $k=1,3$.} \label{pimezzi_0_2}
\end{figure}

\begin{figure}[htbp]
\centering
\makebox{
\includegraphics[width=3in]{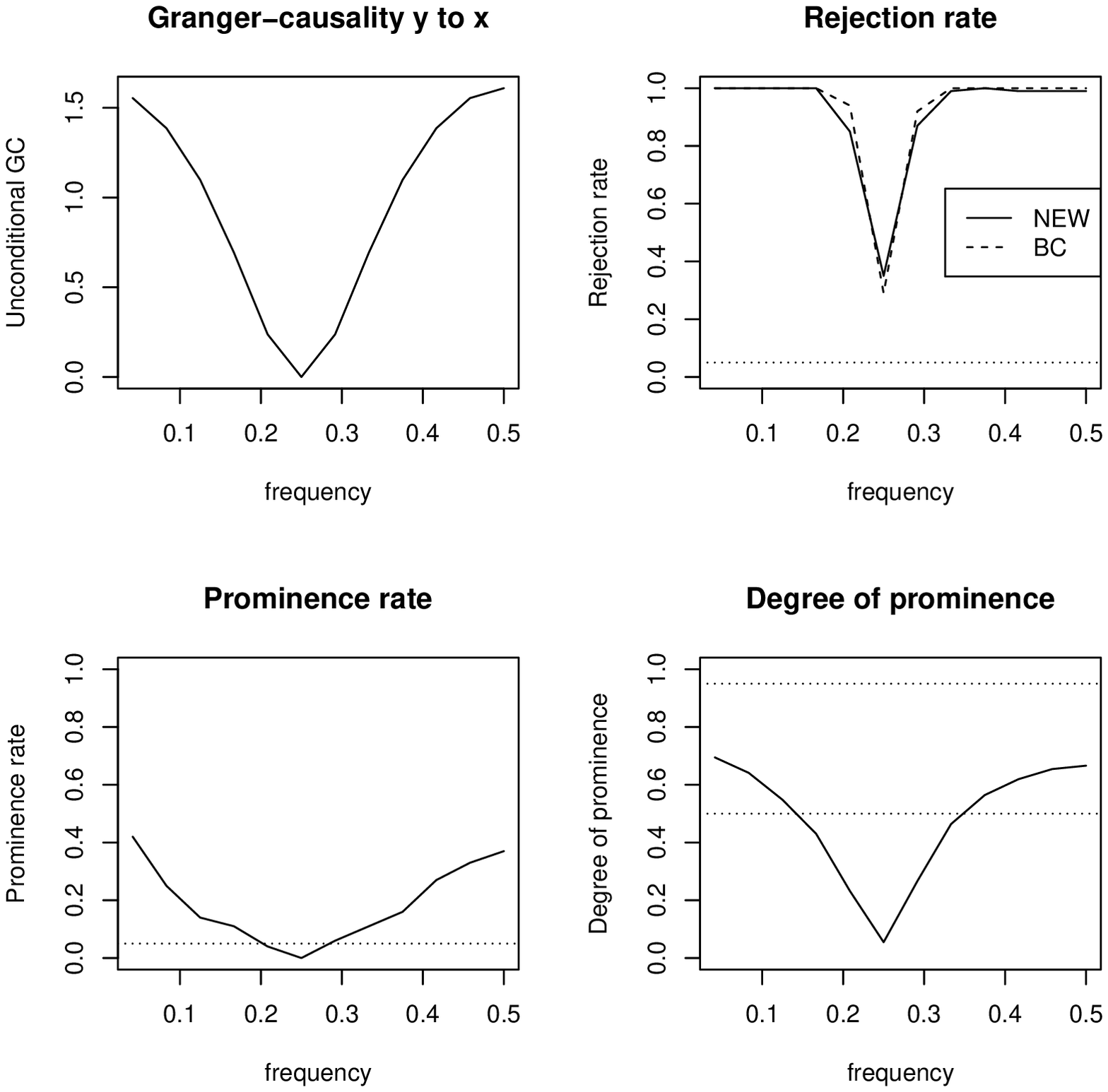}}
\caption{Case with $\omega^{*}=\frac{\pi}{2}$, $A_{k,(22)}=0$, $\matformat{\Sigma}=diag(5,1)$, $k=1,3$.} \label{pimezzi_5}
\end{figure}

If $\omega^{*}=\frac{\pi}{2}$ and $\mathbf{\Sigma}=diag(1,1)$, our rejection rate is $0.8$ at extreme frequencies, and $0.6$ at $\omega^{*}$, resembling the shape of the degree of prominence (Figure \ref{pimezzi_top}). On the contrary, BC test shows a rejection rate of $0.2$ at $\omega^{*}$, and $1$ at extreme frequencies.
Setting $\matformat{\Sigma}=\matformat{diag}(0.2,1)$, the rejection rate of BC test ranges from $0.7$ to $0.2$, while ours is approximately constant around $0.3$
(Figure \ref{pimezzi_0_2}).
This happens because the magnitude of $X$ is much smaller than the one of $Y$, such that $X$ is close to a null process,
and the underlying causality is small and detected as constant across frequencies. 
Setting $\matformat{\Sigma}=diag(5,1)$ (Figure \ref{pimezzi_5}), the rejection rate of both tests stands around $1$, except from a value of $0.3$ at $\omega^{*}$.
This occurs because the magnitude of $Y$ is much smaller than the one of $X$, such that any non-null causality is detected as maximally prominent.


Setting $A_{k,(22)}$, $k=1,3$, to $0.25$ and $0.5$ 
equals to increase the magnitude of the VAR roots until the limit value of $1$ (non-stationary case).
In that case, we observe that the range of our rejection rate increases accordingly to the degree of prominence, achieving $1$ in the non-stationary case.
Our competitor detects much better the null causality, while it is less able, as expected, to catch the shape of the degree of prominence across frequencies.





If we set, as in \cite{breitung2006testing}, $\omega^{*}=0$, $\omega^{*}=\frac{\pi}{4}$, $\omega^{*}=\frac{3\pi}{4}$ and $\omega^{*}={\pi}$, we note that our competitor is less precise, as described therein, particularly for the first two cases, because the rejection rate is considerably above $5\%$. Its rejection rate for non-zero causalities is $100\%$, while ours resembles the shape of the degree of prominence, which tends to $0$ for null causalities with particular intensity for the cases $\omega^{*}=0$, $\omega^{*}=1$.

To sum up, the rejection rate of our test depends on three factors:
\begin{itemize}
\item the magnitude of VAR roots, which has the effect to extend the range. 
In general, the rejection rate is perturbed at low frequencies as the process is closer to non-stationarity;
\item the true underlying spectral variability, which in turn depends on the relationship between the magnitude of causality and non-causality coefficients;
\item the condition number of the autocovariance matrices $\matformat{R}_j$, $j\geq 0$, which masks the underlying spectral variability.
\end{itemize}

In Table \ref{bonf} we report the rejection rates of the test on all causalities jointly considered obtained by Bonferroni correction. We note that the test has power approximately $0.05$ in case of no-causality (Case 3), and approximately $1$ in case of non-stationarity (Cases 2 and 7).
\begin{table}
  \centering
\begin{tabular}{cc}
\hline
Case & Rejection rate \\
\hline
1	& 0.48\\
2	& 0.98\\
3	& 0.05\\
4	& 0.62\\
5	& 0.67\\
6	& 0.16\\
7	& 0.99\\
\hline
\end{tabular}
  \caption{Test on all causalities jointly considered obtained by Bonferroni correction.\label{bonf}}
\end{table}

\section{A Granger-causality analysis of Euro Area GDP, M3 and M1 in the frequency domain}\label{sec3}

While remembering Friedman and Schwartz's general statement (see\\ \cite{friedman2008monetary})
that ``In monetary matters appearances are deceiving: the important relationships are often
precisely the reverse of those that strike the eye'', in this section we study
the co-movements of gross domestic product (GDP) and money stock (M3 and M1 aggregate) in the Euro
Area. We test in the frequency domain both the \textit{direct} link from one variable
to the other one and the \textit{indirect} link with respect to further
explanatory variables like the inflation rate (HICP), the unemployment rate (UN), or the long-term interest rate (LTN).

Published works on this research topic make use of time-domain methods:
some of them use factor modelling \citep{cendejas2014business},
some others use likelihood methods (\cite{andres2006money},
\cite{canova2011does}), or large-dimensional VAR models \citep{giannone2013money}, or VAR models with time-varying parameters \citep{psaradakis2005markov}.
A good review for the pre-Euro period may be found in \citet{hayo1999money}, which explored the relationship
between business cycle and money stock in EU countries via a Granger-causality analysis in the time domain, exactly as
\cite{tsukuda1998granger} did for the Japanese economy.

On the contrary, we apply the inferential framework for GC in the frequency domain developed in Section \ref{sec:G}.
Differently from \cite{breitung2006testing}, which tests the nullity of Granger-causalities at each frequency,
our test is able to discern prominent causalities in comparison to others.
In this way, we provide explicit inference on unconditional and conditional GC.
HICP, UN and LTN are used as conditioning variables, with the aim to discount for the mediating power of each of the
three variables with respect to the relationship between output and money supply.
The same approach also allows us to compare unconditional and conditional GC relative to the same directional link.

\subsection{Data preparation}

We have considered the time series of GDP at market price
in the Euro Area (chain linked volumes in Euro) and
the monetary aggregate M3 and M1 (outstanding amount of loans to the whole economy excluded
the monetary and financial sector, all currencies combined).
M3 is also called ``broad money'', M1 ``narrow money''.

There is not a general consensus on which measure of money supply is the most
appropriate. While the Federal Reserve has officially ceased to publish M3
series since 2006, the M3 index of notional stocks, i.e. the annual growth
rate of the outstanding amount (also called ``base money''), is still used by
the ECB as the official measure of short-term circulating money. For a nice
discussion on the role of M3 as a policy target for central bankers see for
example \citet{alves2007euro}.


Since our goal is to focus on the effect of monetary policy on output, we
restrict our analysis to the period 1999-2017, when the ECB has taken
actual decisions on the Euro Area.
Monthly series (all but GDP) are made quarterly by averaging.
We can thus denote our series by
${GDP}_t$, ${M3}_t$, ${M1}_t$, ${HICP}_t$, ${UN}_t$, ${LTN}_t$,
where $t=1,\ldots,56$ (there are $56$ quarters from Winter 2001 to Autumn 2014).
The data are drawn from the ECB Real Time Research database
where national figures are aggregated according to a changing composition of the Euro Area across time
(see \cite{giannone2012area}).
We refer to \verb"https://www.ecb.europa.eu/stats" and \cite{MFI} for technical and computational details.

\begin{figure}[htbp]
\centering
\makebox{
\includegraphics[width=3in]{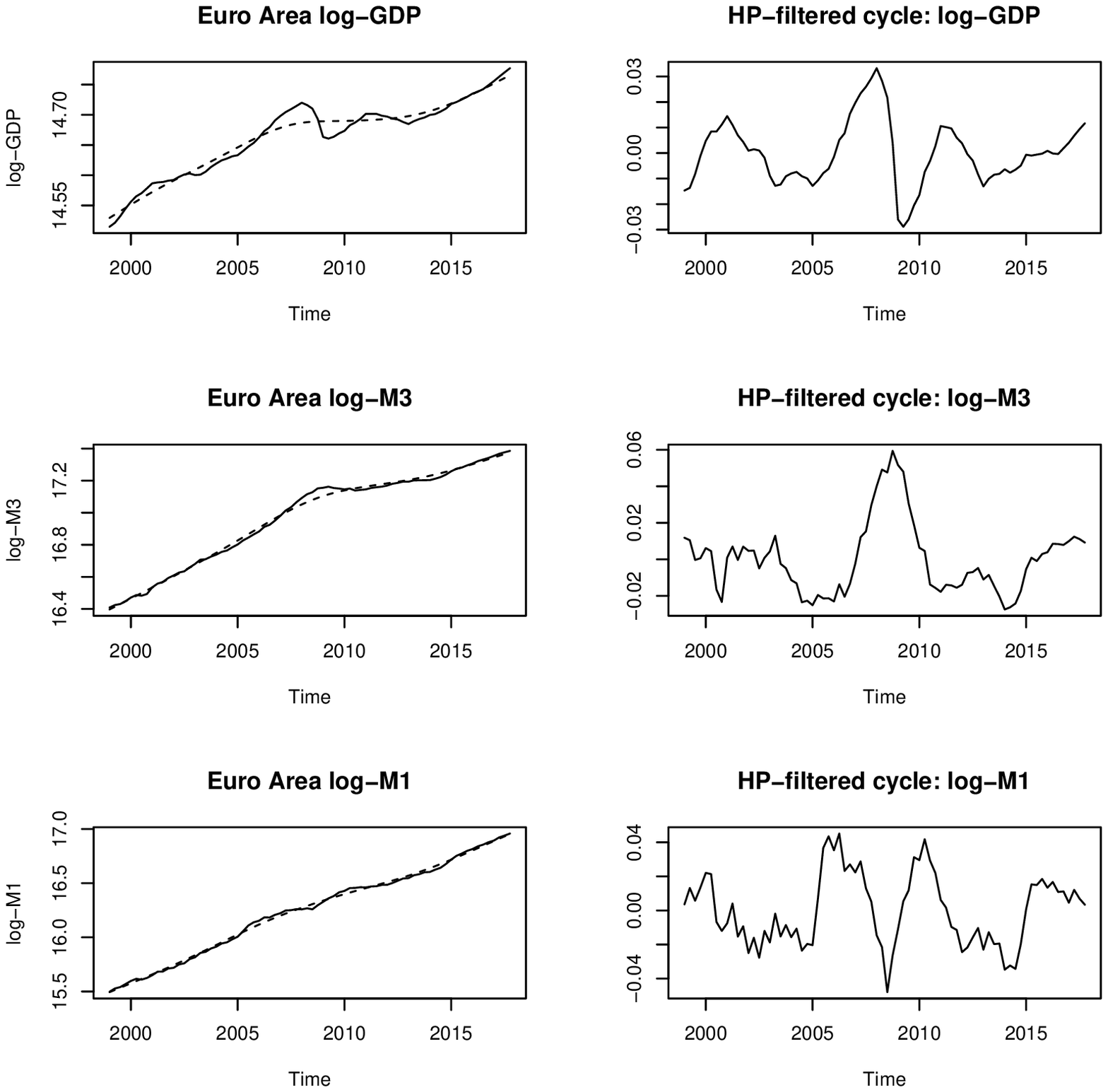}}
\caption{GDP, M3 and M1 in logs - Euro Area. In dashed the extracted trend.} \label{orig_plot_top}
\centering
\makebox{
\includegraphics[width=3in]{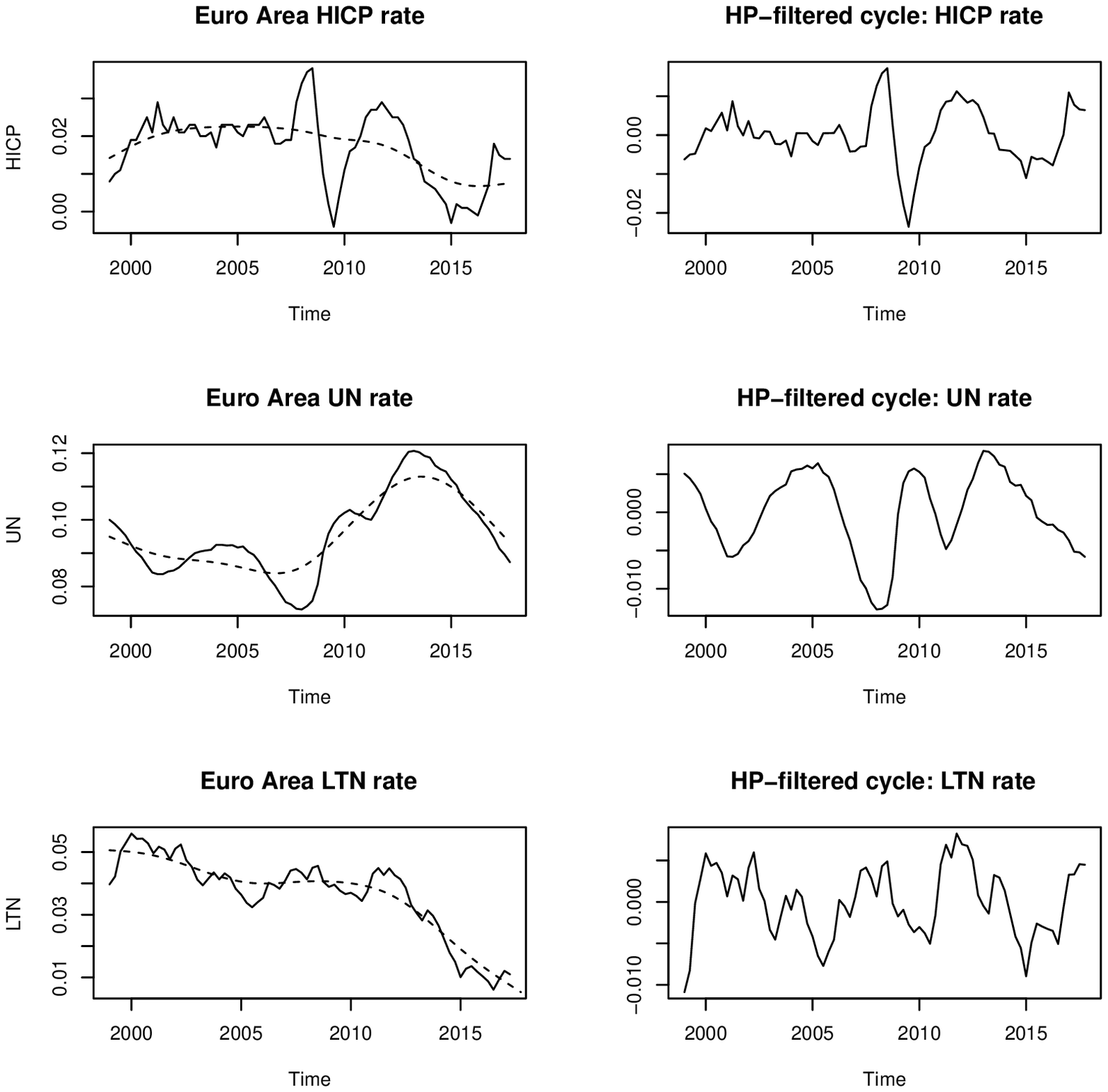}}
\caption{HICP, UN, LTN rates - Euro Area. In dashed the extracted trend.} \label{orig_plot_top_2}
\end{figure}

According to Dickey-Fuller test, the logarithmic transform of $GDP_t$, $M3_t$, $M1_t$ are non-stationary,
as well as the three conditioning variables $HICP_t$, $UN_t$ and $LTN_t$.
Therefore, following \cite{friedman1975money},
we pass all series
by Hodrick-Prescott filter \citep{hodrick1997postwar},
with the canonical value of $\lambda=1600$, in order to remove any trend and to extract cyclical components.
We do not use Baxter-King filter \citep{baxter1999measuring}, as suggested in \cite{belongia2016money},
because we have not enough end of sample data. Cycle extraction is performed via the R package ``mFilter''.

\begin{figure}[htbp]
\centering
\makebox{
\includegraphics[width=3in]{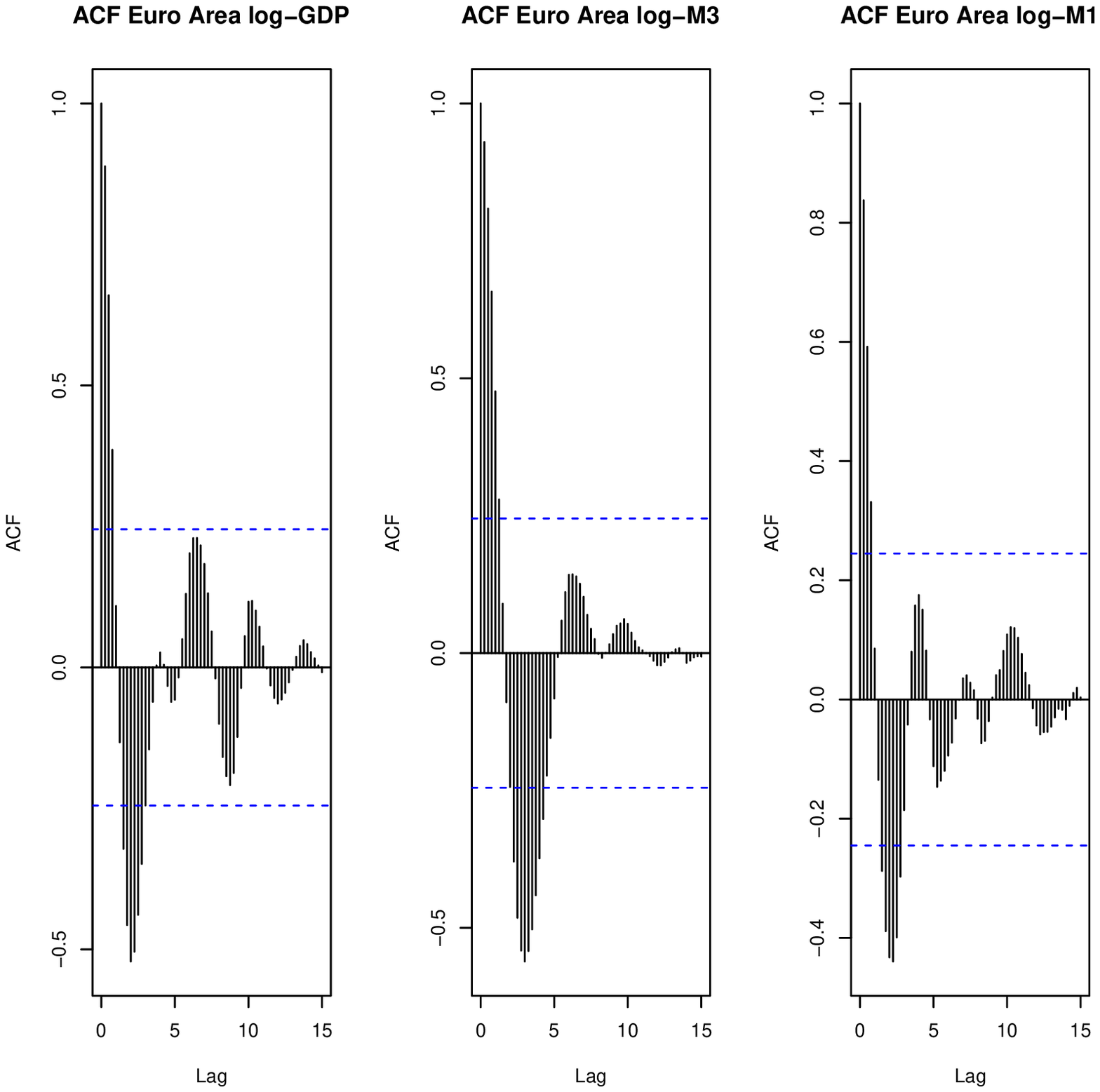}}
\caption{ACF of GDP, M3 and M1 in logs - Euro Area.} \label{acf_log}
\centering
\makebox{
\includegraphics[width=3in]{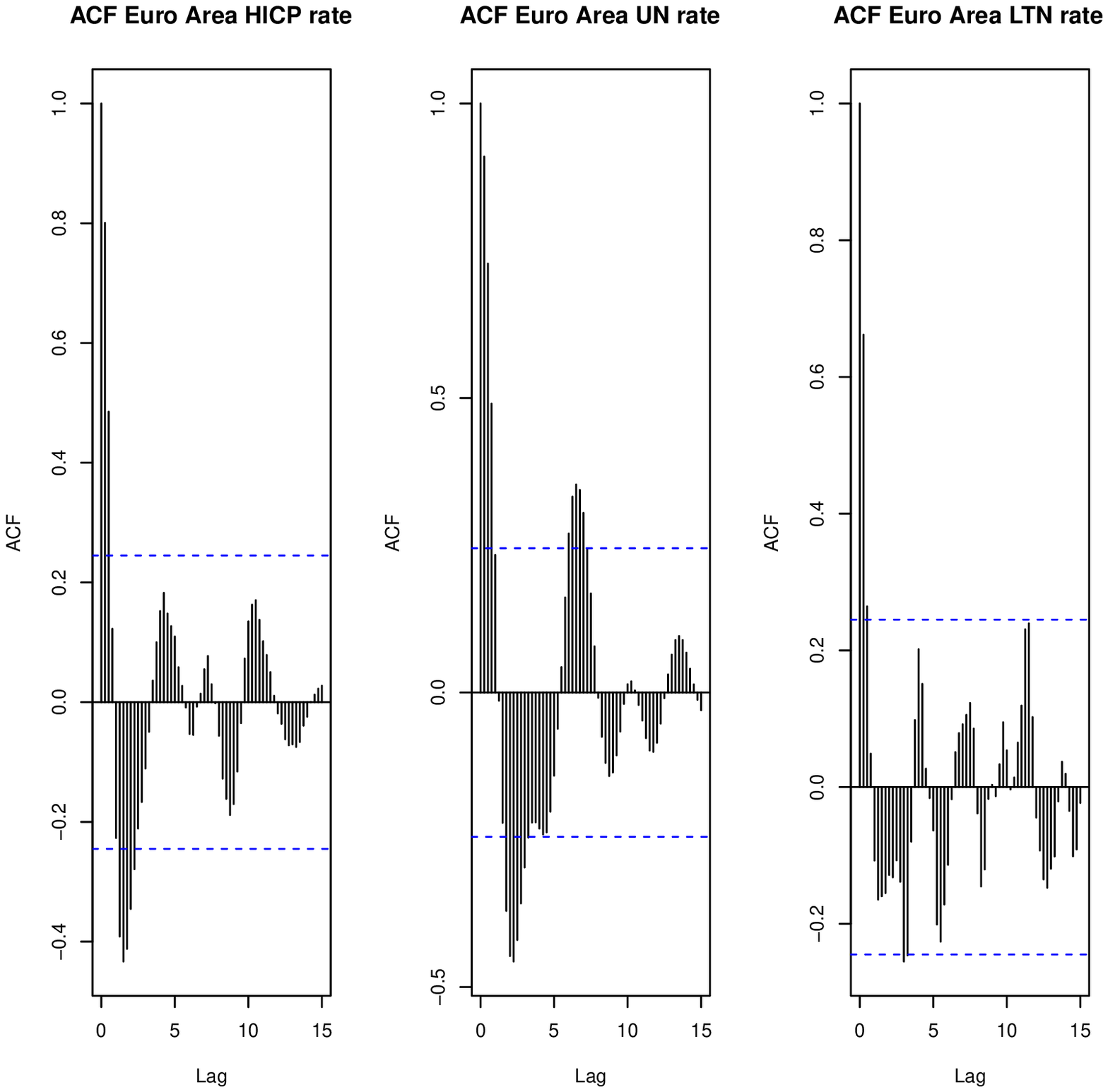}}
\caption{ACF of HICP, UN, LTN rates - Euro Area.} \label{acf_diff}
\end{figure}

\begin{figure}[htbp]
\centering
\makebox{
\includegraphics[width=3in]{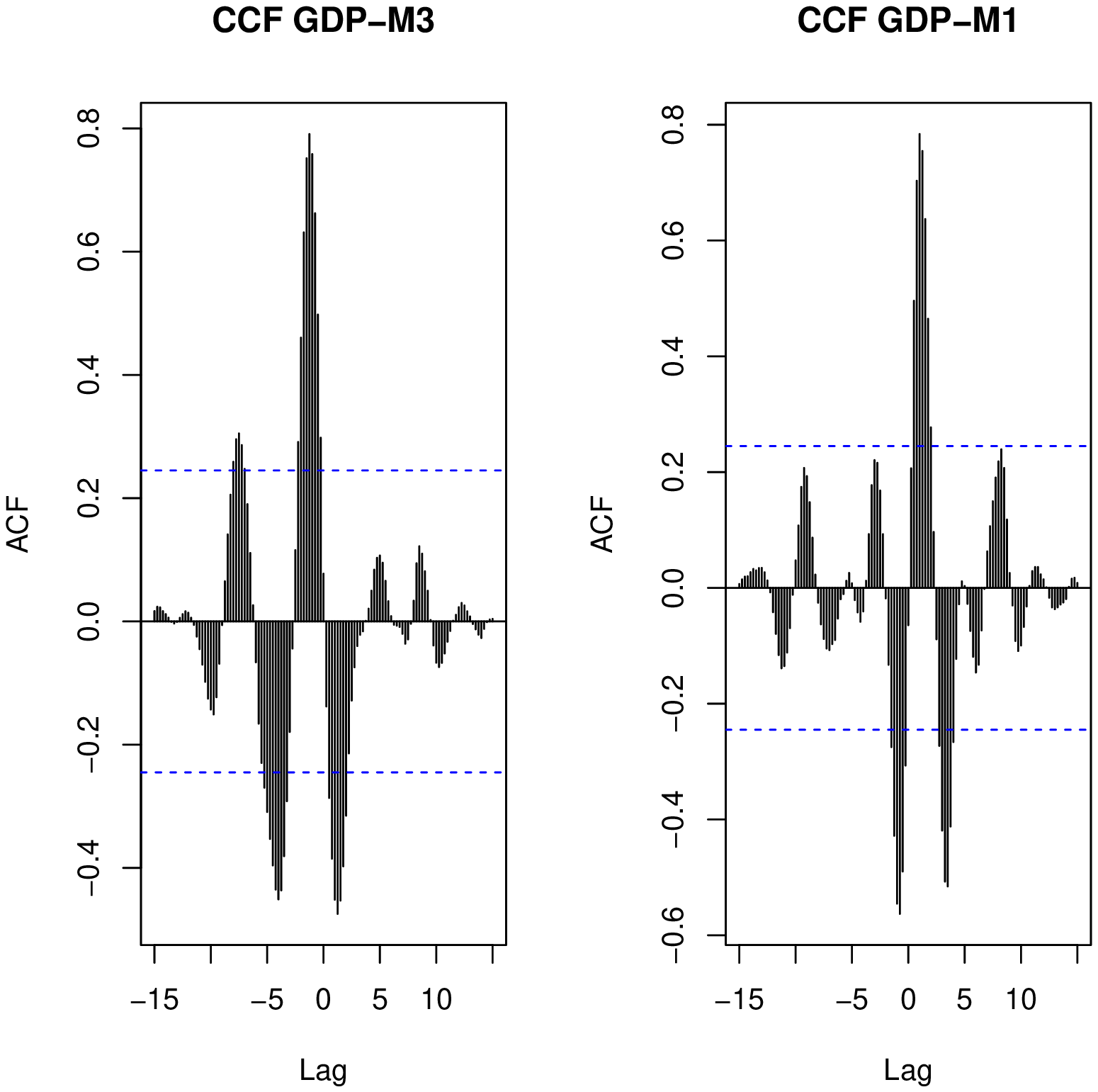}}
\caption{CCF GDP-M3 and GDP-M1 - Euro Area.} \label{ccf_log}
\end{figure}

Figures \ref{orig_plot_top} and \ref{orig_plot_top_2} contain the plots of $GDP_t$, $M3_t$, $M1_t$ and $HICP_t$, $UN_t$, $LTN_t$ respectively.
Left figures contain the original series and the estimated trend, while right figures contain the estimated cycles.
Figures \ref{acf_log} and \ref{acf_diff} show the ACF of the extracted cycles. The patterns are very similar across series: positive for the first 4-5 quarters, negative for all quarters around 2 years and non-significant elsewhere. $UN_t$ shows a rebound for the quarters around $5$ years. $LTN_t$ is no longer significant after $2$ quarters.
Figure \ref{ccf_log} shows the CCF for the couples GDP-M3 and GDP-M1. Their pattern is similar: we have positive correlation around $0$ and
negative correlation at sides around the lag of $2$ years. 

Since our ultimate goal is
to infer about the cause-effect relationship of money stock and economic
output, we test at each frequency the equality between
Granger-causality spectra and the median GC across frequencies,
both unconditional and conditional on the inflation rate, the unemployment rate and the long-term interest rate.
In this way, we can display the relevant cycles in the causality structure
of the relationship from GDP to M3 (M1) and viceversa.
Due to the use of Fast Fourier Transform, the frequencies used are the following:
$f_i=\frac{i}{80}, i=1,\ldots,40$, because $T=76$.
The frequency range is re-scaled to $[0,2]$ for the
quarterly frequency of our series.

Relevant VAR models, estimated including an intercept
by the R package ``vars" \citep{pfaff2008var},
are selected by the Bayesian Information Criterion (BIC), imposing a
maximum of four lags.
BIC is used because we know that BIC is correctly
estimating the unknown number of delays, while AIC may
overestimate it, thus increasing the probability to estimate
non-stationary VAR models. 
In any case, all roots of estimated characteristic polynomials are strictly smaller than one.
In the end, the resulting number of delays is then fixed across the bootstrap inference procedure for each VAR estimation.
The number of bootstrap samples is $1000$.

Note that for computational reasons BC test cannot be computed for $k=1$. Besides, its p-value is constant across frequencies (except the last one) for $k=2$.
BC test requires a large number of delays, while ours works for all values, given that the resulting VAR is stationary and non-singular. Therefore, we can not compare directly our test to BC test on real data, because BC is not useful for all cases with $k\leq2$.


\subsection{Causality results}


We start describing VAR estimates on the couple GDP-M3. Our lag selection procedure chooses $2$ lags.
In the ${GDP}_t$ equation, ${GDP}_{t-1}$ and ${GDP}_{t-2}$ are heavily significant, while ${M3}_{t-1}$ and ${M3}_{t-2}$ slightly are (at $5\%$ and $10\%$ respectively).
This results in a GC spectral shape which is approximately constant across frequencies.
In the ${M3}_t$ equation, $M3_{t-1}$ is heavily significant, while $GDP_{t-2}$ is at $10\%$. The corresponding GC shape is prominent at low frequencies.

Concerning the couple GDP-M1, our VAR lag selection procedure chooses $2$ lags.
In the ${GDP}_t$ equation, ${GDP}_{t-1}$, ${GDP}_{t-2}$ and ${M1}_{t-2}$ are heavily significant. The related unconditional GC shape is prominent at low frequencies only.
In the ${M1}_t$ equation, only ${M1}_{t-1}$ is heavily significant, while $GDP_{t-1}$ has a p-value of $12\%$.
The resulting GC spectral shape is thus prominent only at very low frequencies.

\begin{figure}[htbp]
\centering
\makebox{
\includegraphics[angle=270,width=3in]{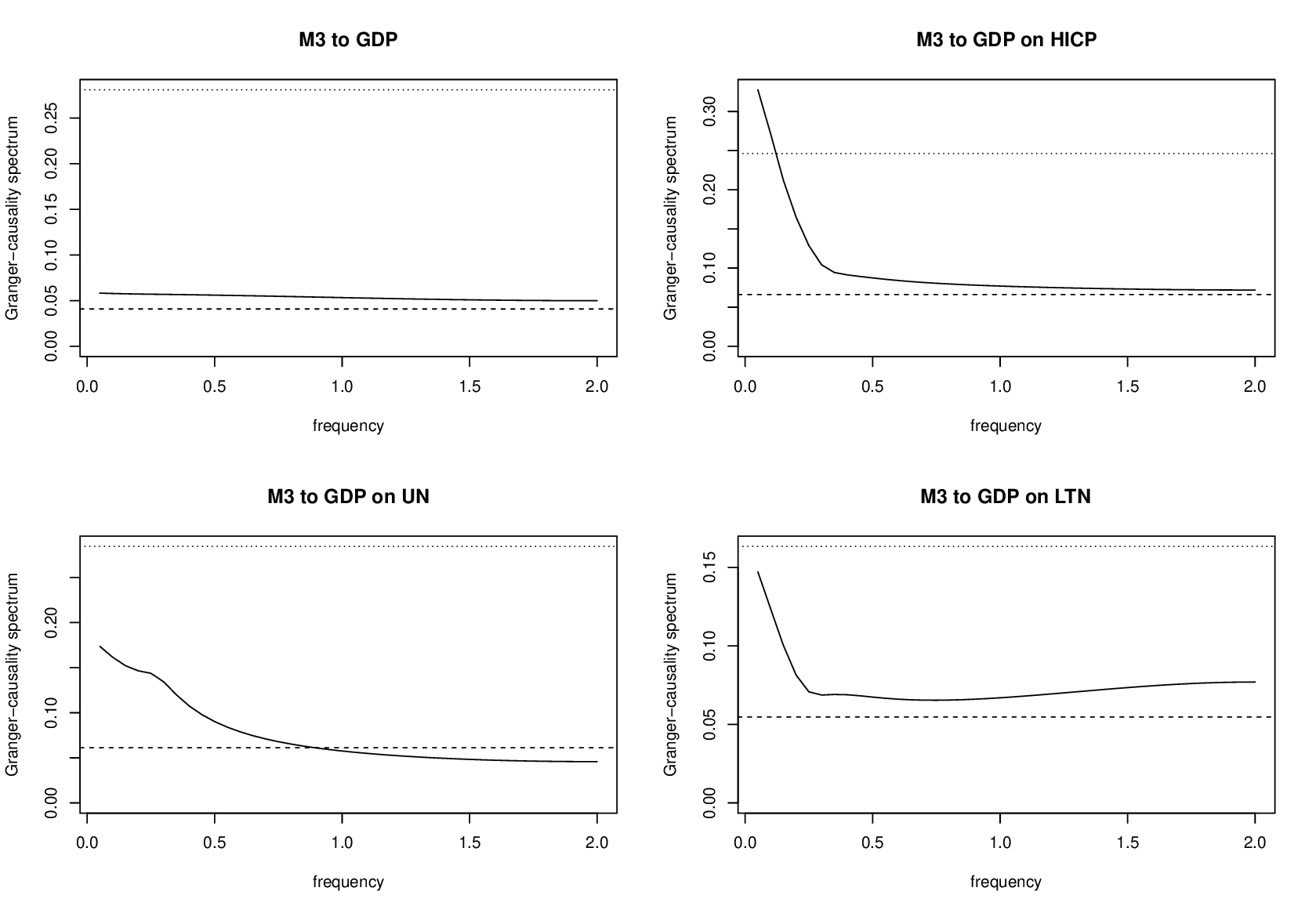}}
\caption{GC spectra M3 to GDP} \label{m3_to_gdp}
\centering
\makebox{
\includegraphics[angle=270,width=3in]{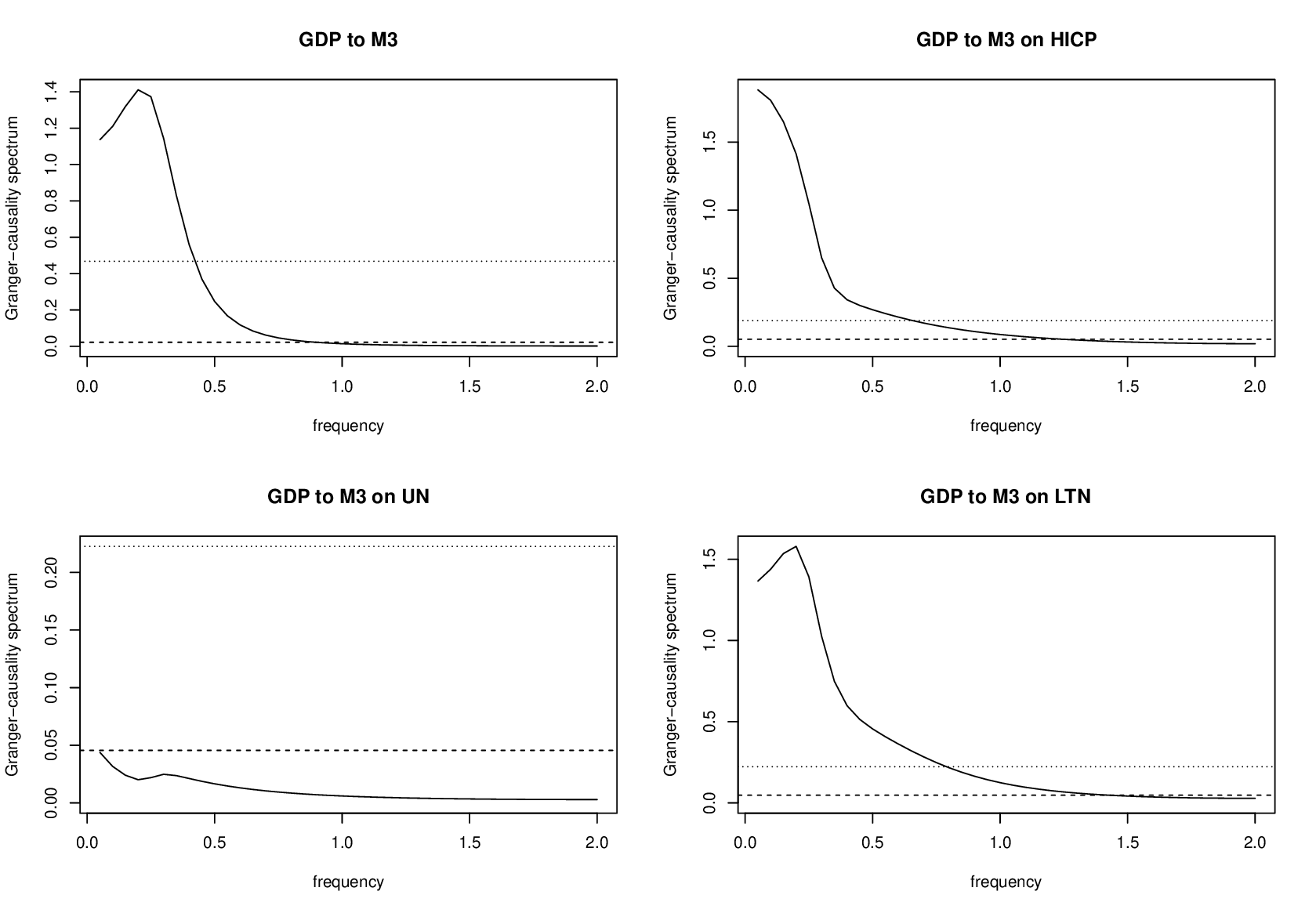}}
\caption{GC spectra GDP to M3} \label{gdp_to_m3}
\end{figure}

\begin{figure}[htbp]
\centering
\makebox{
\includegraphics[angle=270,width=3in]{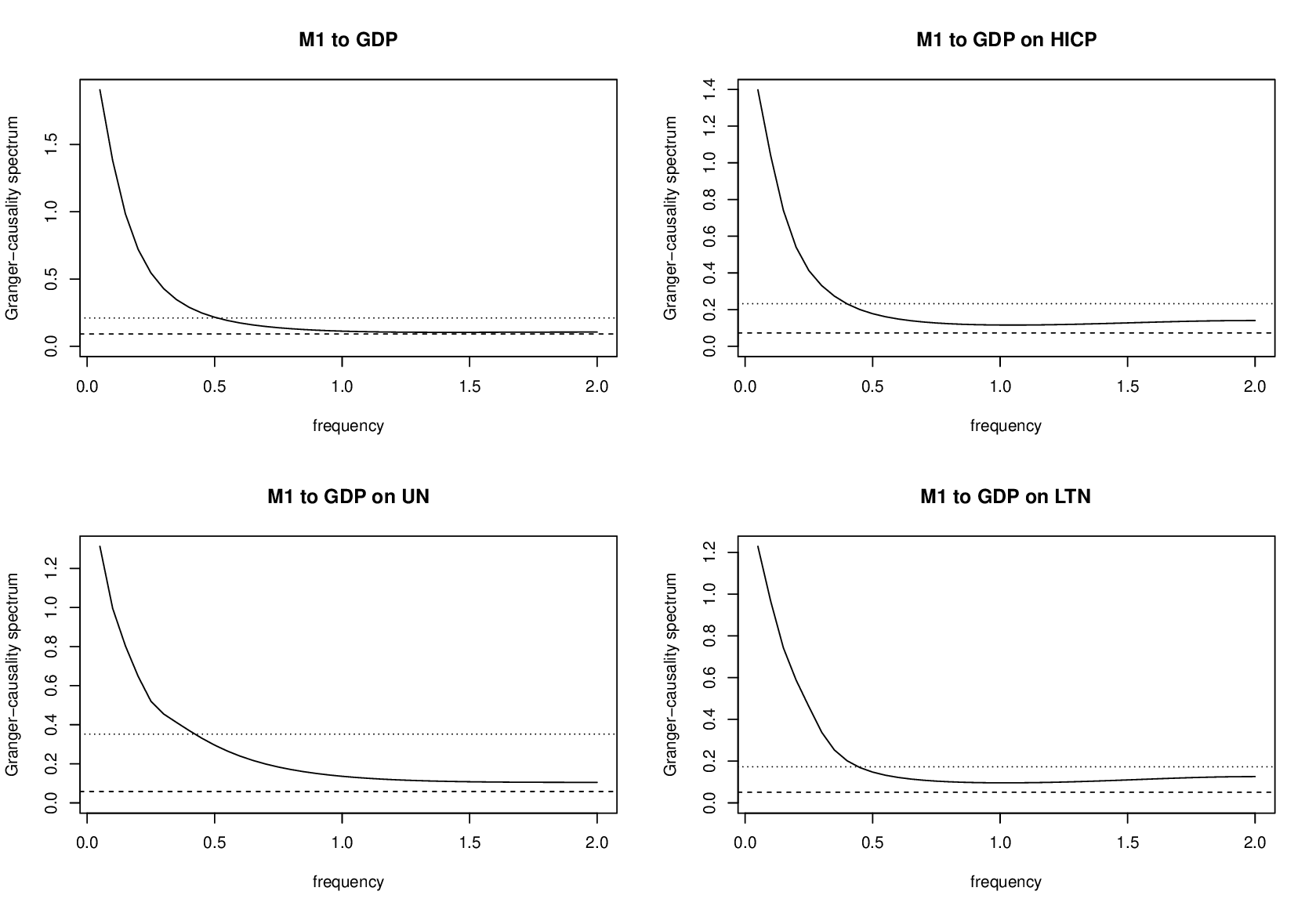}}
\caption{GC spectra M1 to GDP} \label{m1_to_gdp}
\centering
\makebox{
\includegraphics[angle=270,width=3in]{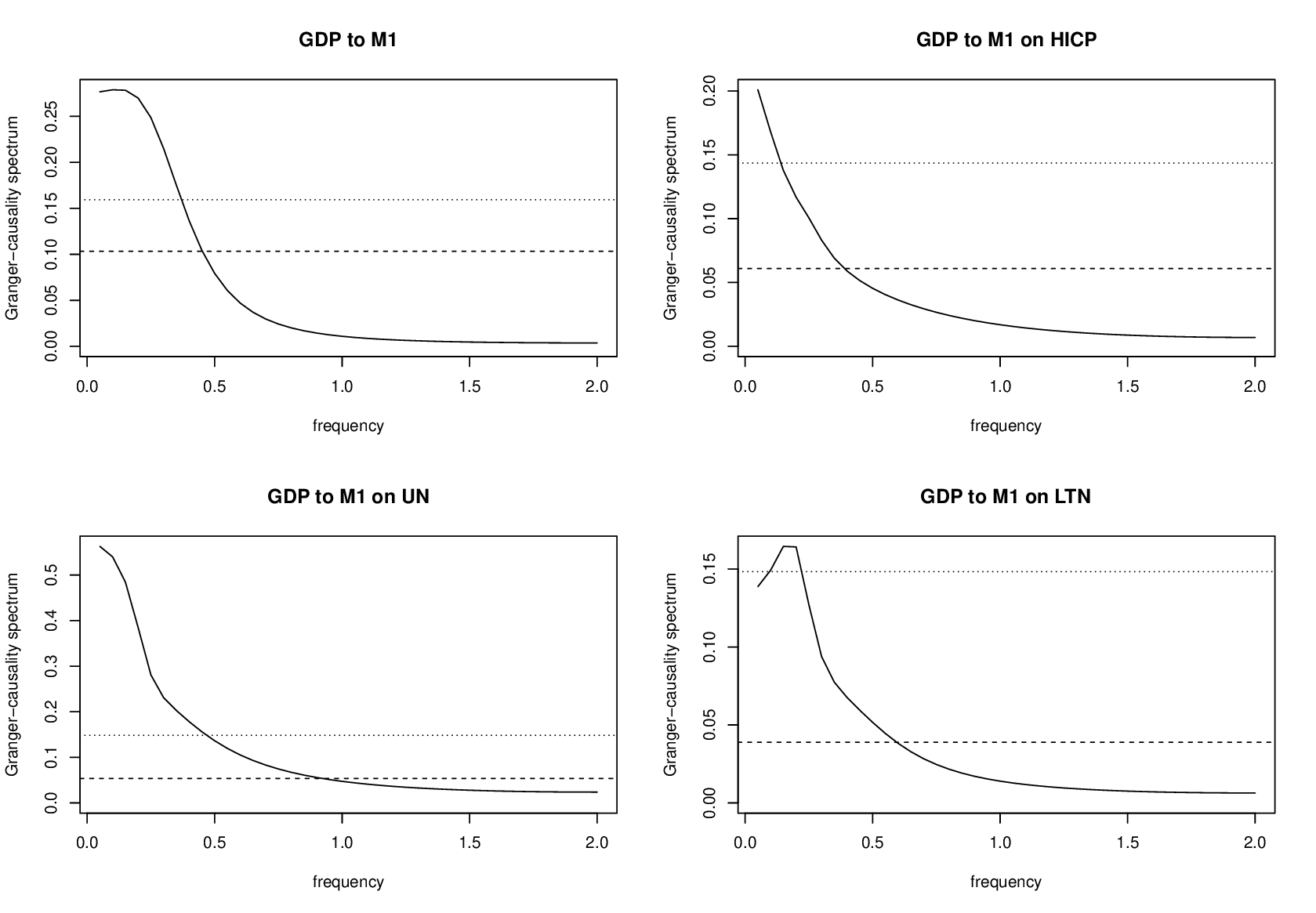}}
\caption{GC spectra GDP to M1} \label{gdp_to_m1}
\end{figure}

In Figures \ref{m3_to_gdp} and \ref{gdp_to_m3}, unconditional and conditional GC spectra from M3 to GDP and viceversa are reported.
The same spectra from M1 to GDP and viceversa are reported in Figures \ref{m1_to_gdp} and \ref{gdp_to_m1} respectively.
In dashed the bootstrap threshold at 5\% is outlined. In dotted, the same threshold for the overall test obtained by Bonferroni correction is depicted.

We first comment conditional GC spectra for the couple GDP-M3.
Conditioning on HICP, the level of significance of ${M3}_{t-1}$ and ${M3}_{t-2}$ is increased in the ${GDP}_t$ equation.
This results in a GC decreasing across frequencies and prominent across the entire frequency range.
In the $M3_t$ equation, the level of significance of ${GDP}_{t-2}$ increases to $5\%$. As a result, GC is prominent until the period of $1$ year.
Conditioning on UN, in the ${GDP}_t$ equation the level of significance of ${M3}_{t-1}$ is $5\%$ while ${M3}_{t-2}$ is no longer significant.
This results in a GC prominent only across the left half of the frequency range.
In the $M3_t$ equation, ${GDP}_{t-2}$ is no longer significant, resulting in a non-prominent GC everywhere. 
Conditioning on LTN, the level of significance is $5\%$ for ${M3}_{t-1}$ and $10\%$ for ${M3}_{t-2}$ in the $GDP_t$ equation.
The corresponding GC is prominent across the entire frequency range.
In the $M3_t$ equation, ${GDP}_{t-2}$ is significant at $5\%$, causing again GC to be prominent everywhere.

We now comment conditional GC spectra for the couple GDP-M1.
Conditioning on HICP, in the $GDP_t$ equation ${M1}_{t-2}$ is still heavily significant. The spectral shape is almost the same as the unconditional one.
In the $M1_t$ equation, the level of significance is quite smaller, so that the only prominent causality is at the lowest frequency.
Conditioning on UN,
${M1}_{t-2}$ is still heavily significant in the $GDP_t$ equation. The spectral shape is very close to the unconditional one (even if slightly weaker).
In the $M1_t$ equation, $GDP_{t-1}$ has a p-value of $20\%$ and the related GC shape is close to the unconditional one.
Conditioning on LTN, ${M1}_{t-2}$ is still significant at $1\%$ in the $GDP_t$ equation, causing GC shape to be almost the same as the one conditioning on UN.
In the $M1_t$ equation, $GDP_{t-1}$ has a p-value of $26\%$. As a consequence, we observe prominence only at the lowest frequency. 

Concerning the overall test on all causalities, we observe the absence of any significance in four cases out of sixteen: the GC spectra from M3 to GDP, unconditional and conditional both on UN and LTN, and the GC spectrum from GDP to M3 conditional on UN. We remark that this test is conservative in nature: however, it allows to adequately
contextualize the significance of individual tests.

\begin{figure}[htbp]
\centering
\makebox{
\includegraphics[angle=270,width=3in]{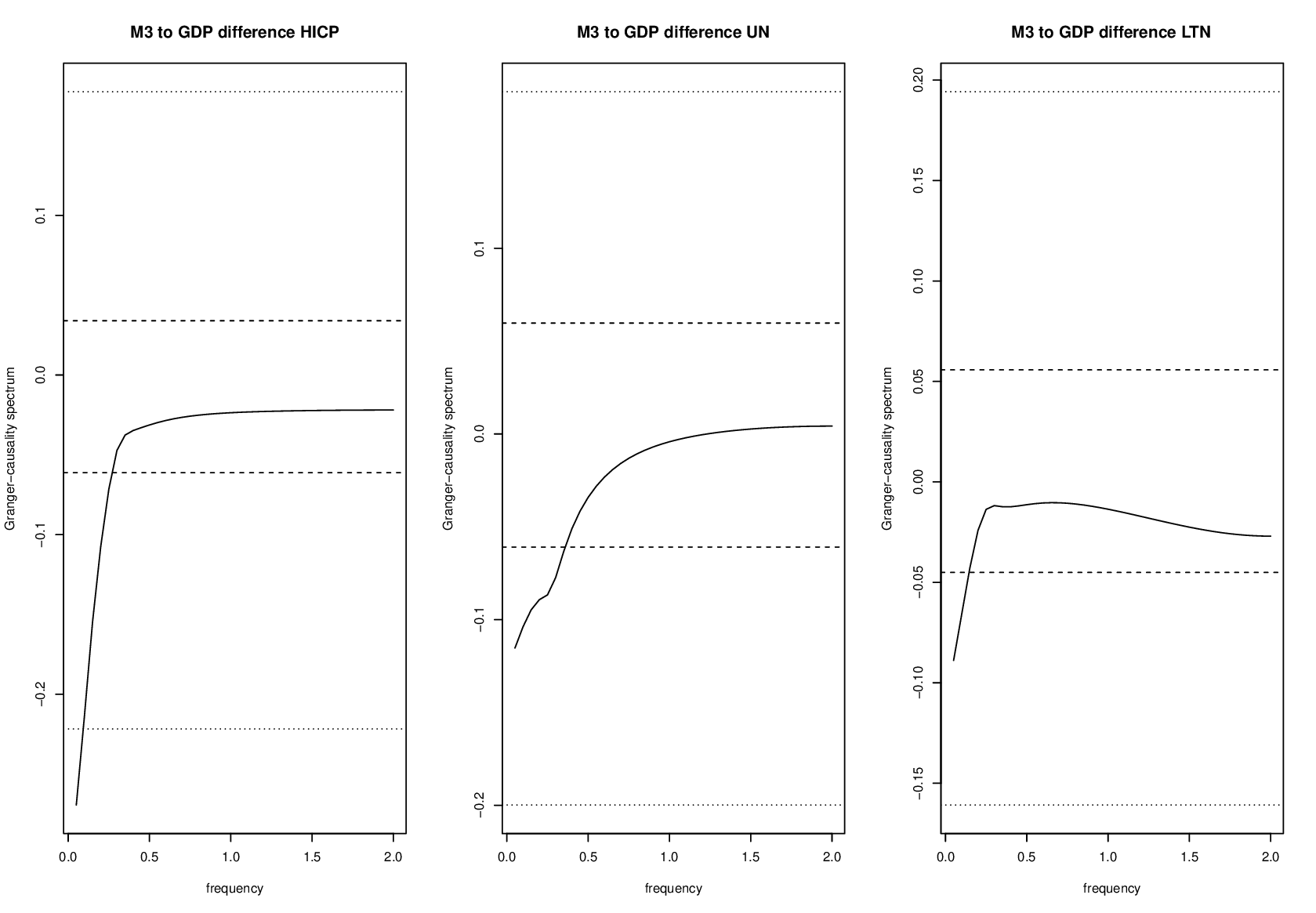}}
\caption{GC spectral differences M3 to GDP} \label{m3_to_gdp_diff}
\centering
\makebox{
\includegraphics[angle=270,width=3in]{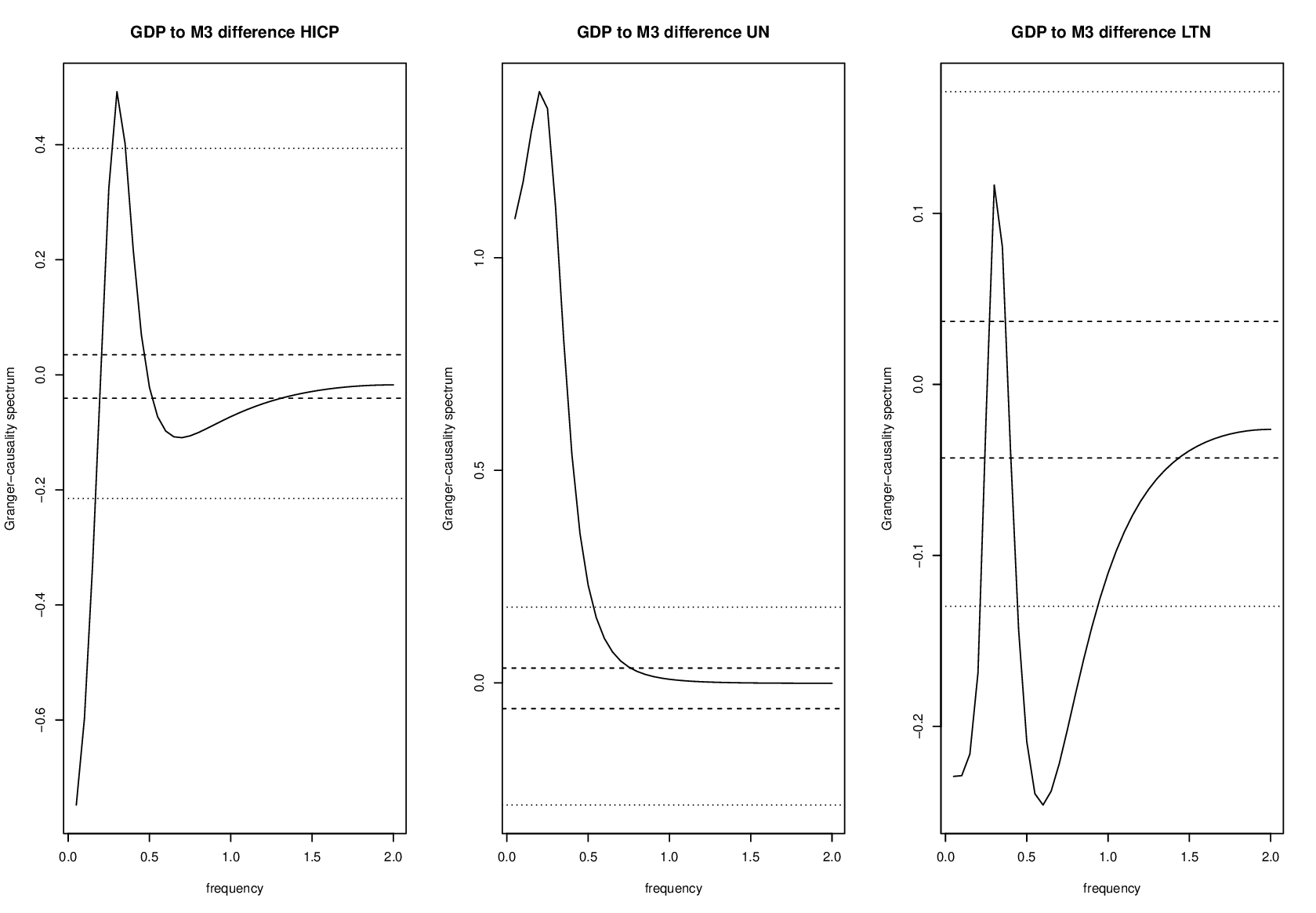}}
\caption{GC spectral differences GDP to M3} \label{gdp_to_m3_diff}
\end{figure}

\begin{figure}[htbp]
\centering
\makebox{
\includegraphics[angle=270,width=3in]{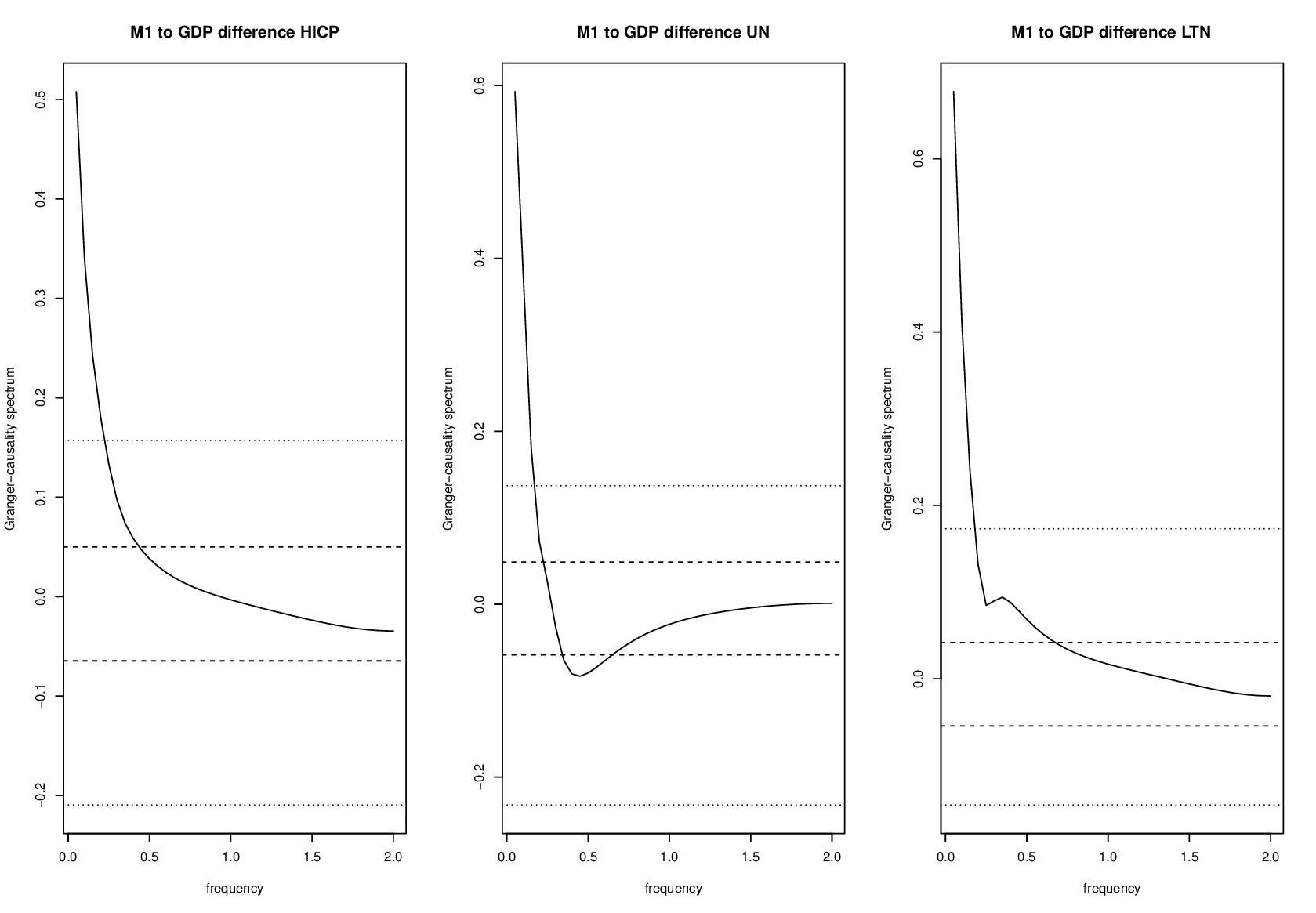}}
\caption{GC spectral differences M1 to GDP} \label{m1_to_gdp_diff}
\centering
\makebox{
\includegraphics[angle=270,width=3in]{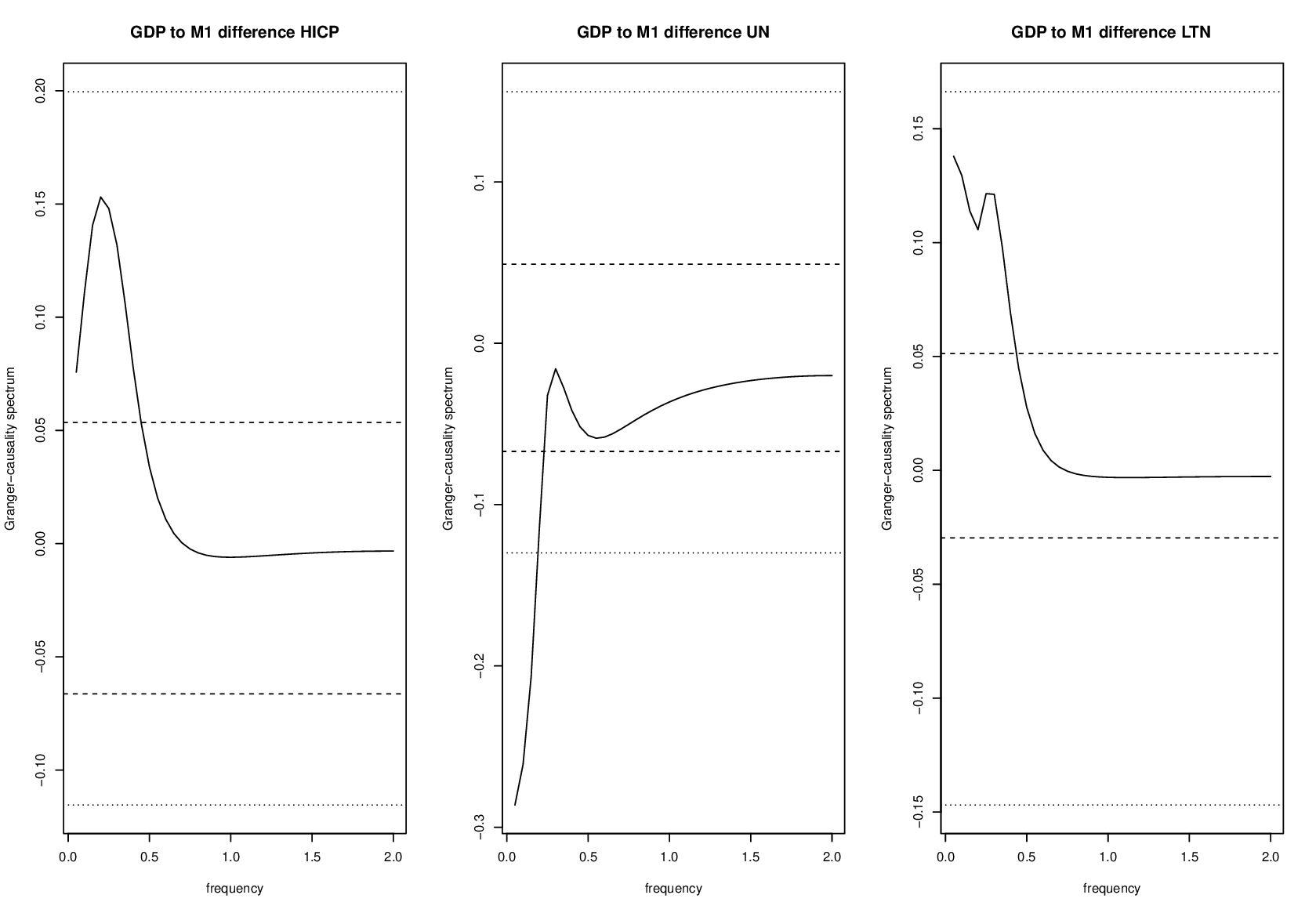}}
\caption{GC spectral differences GDP to M1} \label{gdp_to_m1_diff}
\end{figure}

Finally, GC spectral differences are reported in Figures \ref{m3_to_gdp_diff}, \ref{gdp_to_m3_diff} for the couple $M3-GDP$,
in Figures \ref{m1_to_gdp_diff}, \ref{gdp_to_m1_diff} for the couple $M1-GDP$ .
From M3 to GDP, we only observe a remarkable amplification power of HICP at low frequencies. UN and LTN show no-causality influence even according to the overall test.
From GDP to M3, HICP and LTN show amplification power at the lowest frequency and annihilation power around the period of 2 years.
On the contrary, UN amplifies the causal relationship across the left quarter of the frequency range.
From M1 to GDP, the three conditioning variables show annihilation power at very low frequencies.
From GDP to M1, UN is observed to amplify the link at low frequencies, while the impact of HICP and LTN is not remarkable even according to the overall test.

\subsection{A summary of results}
To sum up, the causal relationship from M3 to GDP is prominent only conditionally on HICP, which appears to be an amplifier, at low frequencies.
We can say that conditionally on HICP the low frequency components of M3 appear to be good predictors of the same components of GDP one step ahead.
The causal relationship from GDP to M3 is also present at low frequencies, except if we condition on UN, which shows a strong annihilation power.
On the contrary, the causality from M1 to GDP is prominent, both unconditionally and conditionally on HICP, UN and LTN, at all frequencies.
The three explanatory variables show a remarkable annihilation power at low frequencies.
In the end, the causality from GDP to M1 appears strong at low frequencies only. The impact of HICP and LTN on the link can be assumed to be non-remarkable,
while UN shows a remarkable amplification power at very low frequencies.

\section{Conclusions and discussion}

The motivating application of this paper was the study of the time relationships between M3
(M1) aggregate and GDP in the Euro Area.
Our ultimate goal was to determine how M3 (M1) affects (or is affected by)
economic output, both \textit{tout court} and taking into account their
relationship with monetary (inflation rate), economic (unemployment rate)
or financial (interest rate) variables.

Granger-causality unconditional spectrum analysis turned out
to be a very effective tool to find out the most relevant time delays in the
reciprocal dynamics of two variables. This is due to the fact that, by this
frequency-domain tool, we can capture all time delays simultaneously
(synthesis power). We can also take into account the latent relationship with
some other variables, computing Granger-causality conditional spectrum.

In this context, we have developed a testing procedure which is able to mark up prominent frequencies,
which are frequencies at which the (unconditional or conditional) causalities are
systematically larger than the median causality.
A simulation study has shown that our test can be used as a complementary tool to \cite{breitung2006testing}, since we do not mark significant causalities but causalities particularly prominent with respect to others. In this way, we can disambiguate among significant causalities the most prominent ones.
In the same way, we are also able to compare unconditional and conditional spectra detecting prominent causality differences.

Our test has a general validity, as it only requires the stationary bootstrap of \cite{politis1994stationary} to be consistent on the data generating process of interest under the hypothesis of no-causality.
Therefore, our procedure may find application in different fields than macroeconomics, like neuroscience, meteorology, seismology and finance, among others.
However, monetary economics is a very suitable application field, as the time series of interest often present a rich causality structure, and the need rises to disambiguate among significant cycles the most prominent ones.

From an empirical point of view, we have been able to say that the relationship between money supply
and output is present in the Euro Area across the period 1999-2017. We have provided evidence that M3 (M1) in some cases reacts
to economic shocks, in some others it acts as a policy shock with respect to
economic output.
We have observed that the link between GDP and M1 is much stronger in both directions than the link between GDP and M3.
In particular, the causal relationship from M1 to GDP appears to be prominent at all frequencies,
while the opposite one is prominent at low frequencies only.

In conclusion, we can say that in the
Euro Area money stock cannot be considered an exogenous variable
\textit{tout-court}, since its interrelation with economic output is complex
and also depends on further explanatory variables in a nontrivial way.
Nonetheless, the intensity of the causal link from money to output appears to be stronger than the reverse one.

\section*{Appendix}\label{proofs}

\subsection*{Proof of Theorem \ref{appr_all}}

Let us define the random vector $\vecformat{Z}_t=[X_t, Y_t, W_t]$. We assume that $X_t$, $Y_t$ and $Z_t$ are stochastically independent, which causes the null hypothesis of no-causality to hold. This is like assuming that the distribution function $F_{\vecformat{Z}}$ can be factorized as $F_{X}F_{Y}F_{W}$. In addition, we assume that $X_t$, $Y_t$ and $Z_t$ are strictly stationary.

By \cite{politis1994stationary} (paragraph 4.3) we know that
\begin{equation}\sqrt{T}(r(\hat{F}_{X})-r(F_{X}))=\frac{1}{\sqrt{T}}\sum_{i=1}^T h_F(X_i)+o(\sqrt{T} ||\hat{F}_{X}-F_{X}||),\end{equation}
where $\hat{F}_{X}$ is the empirical density function of $X_t$
and $F_{X}$ is the corresponding true distribution function. The same equation holds for $Y$ and $Z$.

If, for some $d \geq 0$, $E(h_{F_{X}}(X_{1}))^{2+d}<\infty$, and if it holds $\sum_{k} \alpha_{X}(k)^{\frac{d}{2+d}}<\infty$,
then $\frac{1}{\sqrt{T}}\sum_{i=1}^T h_{F_X}(X_i)$ is asymptotically normal with mean $0$ and variance \begin{equation}E(h_{F_{X}}(X_i)^2)+2 \sum_{k=1}^{\infty} cov(h_{F_{X}}(X_1),h_{F_{X}}(X_{1+k}))\label{var_h}.\end{equation}

The same equation holds for $Y$ and $W$ if $E(h_{F_{Y}}(Y_{1}))^{2+d}$, $\alpha_{Y}(k)^{\frac{d}{2+d}}<\infty$, and $E(h_{F_{W}}(W_{1}))^{2+d}$, $\sum_{k} \alpha_{W}(k)^{\frac{d}{2+d}}<\infty$ respectively.

At the same time, if $\sum_{k} k^2 \alpha_{X}(k)^{1/2-\tau}< \infty$, $\sum_{k} k^2 \alpha_{Y}(k)^{1/2-\tau}< \infty$,\\ $\sum_{k} k^2 \alpha_{Z}(k)^{1/2-\tau}< \infty$ for some $0<\tau < 1/2$, the stochastic processes $\hat{F}_{X}-F_{X}$, $\hat{F}_{Y}-F_{Y}$, $\hat{F}_{W}-F_{W}$ converge in supremum norm to a Gaussian process having continuous paths and mean $0$. Therefore, $\sqrt{T}(r(\hat{F})-r(F))$ is asymptotically normal with mean $0$ and variance (\ref{var_h}).

Moreover, for each random variable the distribution of $\sqrt{T}(r(\hat{F})-r(F))$ is approximated via the distribution of $\sqrt{T}(r(\hat{F}^{*})-r(\hat{F}))$, where $\hat{F}^{*}$ is the empirical density function obtained via stationary bootstrap.
This holds because the two distributions converge to the same Gaussian process under previous weak dependence assumptions,
provided that $T^{\frac{1}{3}} \rightarrow \infty$.

At this point, since we assumed the stochastic independence of $X_t$, $Y_t$ and $Z_t$,
the weak dependence assumptions on
$E(h_F)$, $\alpha_X$ and $\sum_{k} \alpha(k)^{\frac{d}{2+d}}$ are transmitted to the whole process $\vecformat{Z}_t$.
Therefore, for any \textit{Fr\'{e}chet-differentiable} functional $r$ we can write
$$P(r(\hat{F}_{\vecformat{Z}}^{*})-r(F)\leq q_r(1-\alpha))=1-\alpha$$
under the assumption $T^{\frac{1}{3}} \rightarrow \infty$.




\subsection*{R package ``grangers''}
Our paper is complemented by an R package, called ``grangers'', with five functions performing the calculation of unconditional and conditional Granger-causality spectra, bootstrap inference on both, and inference on the difference between them (see \url{https://github.com/MatFar88/grangers}). The package also contains the data used for the analysis, and two functions performing the tests of \cite{breitung2006testing} on unconditional and conditional Granger-causality respectively.

%
%
%
%
%
%

\bibliographystyle{apalike}
\bibliography{bibliotop_aoas}

\end{document}